\documentclass[
    envcountsame
]{llncs}

  \usepackage[T1]{fontenc}
  \usepackage{utf8-math}
  \usepackage{microtype}

  \usepackage[english]{varioref}
  \usepackage[final]{hyperref}
  \hypersetup{unicode,
              breaklinks=true,
              pdfborder={0 0 0},
}
  \usepackage[capitalise]{cleveref}       \crefname{condition}{condition}{conditions}
  \crefname{case}{case}{cases}

  \usepackage{tikz}
  \usetikzlibrary{arrows,arrows.meta,automata,positioning,graphs,bending,quotes, matrix,    calc}
\usepackage{pgffor}

  \usepackage{array}   \usepackage{paralist}
  \usepackage{xspace}

\colorlet{mygreen}{black!50!green}
\colorlet{myred}{black!20!red}

\tikzset{
  >=stealth',
  initial text =,
  node distance = 2.5em,
  micro nodes/.style = {
        >={Stealth[length=2pt, width = 2pt]},
           node distance = 8pt,
     every state/.style = {
       inner sep = 0pt,
       minimum size = 3pt,
       initial distance = 3pt,
            },
    font=\tiny,
        every edge/.style = {thin, draw, shorten >=0.1pt, font=\tiny},
      },
  tiny nodes/.style = {
     node distance = 5.5ex,
     every state/.style = {
       inner sep = 0pt,
       minimum size = 1ex,
       initial distance = 2ex,
            },
    font=\small,
        every edge/.style = {draw, shorten >=1pt, font=\tiny},
      },
  small nodes/.style = {
         every state/.style = {
       inner sep = 0pt,
       minimum size = 2.5ex,
       initial distance = 3ex,
            },
    font=\scriptsize,
        every edge/.style = {
      draw,
      shorten >=1pt,
    },
  },
  loop above left/.style = {
    out=75, in=105, loop
  }
}

  \newcommand{\op}[1]{\ensuremath{\mathsf{#1}}}

  \newcommand{\sym}[1]{\mathsf{#1}}
  \newcommand{\mathsym}[1]{\mathrm{#1}}
  \newcommand{\fov}[1]{\mathsf{#1}}  \newcommand{\fof}[1]{\mathrm{#1}}

  \newcommand{\suc}[1]{\ensuremath{\mathsym{suc}_{\mc{#1}}}}

  \newcommand{\sub}{\mathsym{sub}}

  \newcommand{\org}{\mathrm{org}}
  \newcommand{\Runs}{\runs}
  \newcommand{\runs}{\mathsym{runs}}
  \newcommand{\lastloop}{\mathsym{lastl}}

  \newcommand{\sboxminus}{{\raisebox{-0.3pt}{\scalebox{0.8}{$\boxminus$}}}}   \newcommand{\sboxplus}{{\raisebox{-0.3pt}{\scalebox{0.8}{$\boxplus$}}}}
    \DeclareUnicodeCharacter{22C8}{\mathrel{\bowtie}}   \DeclareUnicodeCharacter{229F}{\sboxminus}   \DeclareUnicodeCharacter{229E}{\sboxplus}

    \newcommand{\symeq}{=}
  \newcommand{\symle}{≤}
  \newcommand{\symge}{≥}
  \newcommand{\symg}{>}
  \newcommand{\syml}{<}

  \DeclareUnicodeCharacter{2A75}{\symeq}   \DeclareUnicodeCharacter{2266}{\symle}   \DeclareUnicodeCharacter{2267}{\symge}   \DeclareUnicodeCharacter{226A}{\syml}   \DeclareUnicodeCharacter{226B}{\symg}
    \newcommand{\LTL}{\op{LTL}\xspace}
    \newcommand{\myLTL}{\op{CLTL}\xspace}

  \newcommand{\fLTL}{\op{fLTL}\xspace}
  \newcommand{\fCTL}{\op{fCTL}\xspace}
  \newcommand{\CTL}{\op{CTL}\xspace}
    
  \newcommand{\CLTL}{\op{CLTL}\xspace}
  \newcommand{\cLTL}{\op{CLTL}\xspace}
  \newcommand{\CCTLpm}{\op{CCTL_±}}
    \newcommand{\CTLstar}{\textsf{CTL{\!*}}\xspace}

  \renewcommand{\U}{\ensuremath{\operatorname{\mathrm{\mathbf{U}}}}\xspace}
  \newcommand{\X}{\ensuremath{\operatorname{\mathrm{\mathbf{X}}}}\xspace}
  \renewcommand{\G}{\ensuremath{\operatorname{\mathrm{\mathbf{G}}}}\xspace}
  
  \newcommand{\Uc}[1]{\ensuremath{\operatorname{\mathbf{U}}_{[#1]}}\xspace}
  \newcommand{\Fc}[1]{\ensuremath{\operatorname{\mathbf{F}}_{[#1]}}\xspace}

  \newcommand{\true}{\xspace\mathsf{true}\xspace}
  \newcommand{\false}{\xspace\mathsf{false}\xspace}
  \newcommand{\ite}{\fof{ite}}

  \newcommand{\FA}{\mathsf{FA}}

  \newcommand{\mc}{\mathcal}

  \newcommand{\QPA}{\textsf{QPA}\xspace}

  \newcommand{\Guards}{ℭ}

  \newcommand{\NP}{\textrm{NP}\xspace}
  \newcommand{\PSpace}{\textsc{PSpace}\xspace}
  \newcommand{\NExp}{\textsc{NExp}\xspace}

\newcounter{resumeenum}

\begin{document}

\title{Flat Model Checking for Counting LTL Using Quantifier-Free Presburger Arithmetic}

\author{}\institute{
\parbox[t]{18em}{\centering {\normalsize{Normann Decker}} \\[1ex] ISP, Universität zu
Lübeck, Germany \\ \email{decker@isp.uni-luebeck.de}}\parbox[t]{18em}{\centering
{\normalsize{ Anton Pirogov}}\thanks{This author is supported by the German research council (DFG) Research Training Group 2236 UnRAVeL}\\[1ex] RWTH Aachen University, Germany \\ \email{pirogov@cs.rwth-aachen.de}}
\vspace{0.3cm}
}

\maketitle

\begin{abstract}
  This paper presents an approximation approach to verifying counter systems with respect to properties formulated in an expressive counting extension of linear temporal logic.
  It can express, e.g., that the number of acknowledgements never
  exceeds the number of requests to a service, by counting specific positions along a run and imposing
  arithmetic constraints.
  The addressed problem is undecidable and therefore solved on flat under-approximations of a system.
  This provides a flexibly adjustable trade-off between exhaustiveness and computational effort, similar to bounded model checking.
  Recent techniques and results for model-checking frequency properties over flat Kripke structures are lifted  and employed to construct a parametrised encoding of the (approximated) problem in quantifier-free Presburger arithmetic.
  A prototype implementation based on the z3 SMT solver demonstrates the effectiveness of the approach based on problems from the RERS Challange.
  \end{abstract}

\section{Introduction}

Counting is a fundamental principle in the theory of computation and well-established in the study and verification of infinite-state systems.
The concept is ubiquitous in programming, and counting mechanisms are a natural notion of quantitative measurement in specification formalisms.
For example, they are useful for expressing constraints such as ``the number of acknowledgements never exceeds the number of requests'' or ``the relative error frequency stays below some threshold''.
An established and well-studied framework for correctness specification is linear temporal logic (\LTL)~\cite{DBLP:conf/focs/Pnueli77}.
Therefore, various counting extensions were proposed~\cite{DBLP:conf/lics/BouajjaniEH95,DBLP:conf/time/LaroussinieMP10,DBLP:conf/tase/BolligDL12,DBLP:conf/concur/DeckerHLST17} that allow for imposing constraints on the number of positions along a run that satisfy some property.
These extensions target different kinds of system models, and vary in the type of events that can be counted and the constraints that can be expressed.

This paper is concerned with verifying properties expressed in the counting temporal logic \myLTL.
This extension of \LTL features a generalised \emph{temporal until} operator $\Uc{.}$ for evaluating a counting constraint within its scope.
For example, consider the property that between two system resets, two events $e_1$ and $e_2$ (say, related sensor events) should be correlated linearly.
The \myLTL formula
\[
  \G (¬\textit{reset}~ \Uc{2e_1-e_2≥-10} \textit{reset})
\]
would specify that there are not more than twice as many occurrences of $e_1$ than there are of $e_2$, with an absolute margin of $10$.
Notice that this property is not regular.
The events $e_1$ and $e_2$ may be atomic or again characterised by some temporal (counting) property.
The definition used\footnote{To avoid cluttered notation when respecting various existing naming schemes, the denotation \myLTL is reused, despite semantic differences.}
here extends that of~\cite{DBLP:conf/time/LaroussinieMP10} by admitting not only natural but arbitrary integer coefficients in constraints.
Without this extension, the logic was shown to be more concise but not more expressive than \LTL.
Moreover, in the present work, \myLTL is interpreted over \emph{counter systems} instead of Kripke
structures and allows for imposing arithmetic constraints also on (linear combinations of) the counter values, similar to the formalisms considered in~\cite{icalp-Cerans94,DBLP:conf/csl/ComonC00,DBLP:journals/iandc/DemriD07}.

Towards making the extended features of this specification language available for program verification, we propose an approach to the \emph{existential model-checking problem} of \myLTL over counter systems, i.e.\ deciding for some counter system whether it admits a run satisfying a given formula.
Both system model and logic are very powerful, and the problem is undecidable.
However, we avoid the often made compromise of recovering decidability by means of essential restrictions to the specification language.
Instead, we use an approximation scheme based on an extension of recent work~\cite{DBLP:conf/concur/DeckerHLST17} that has laid the theoretical basis for a decision procedure in the special case of structures that are \emph{flat}.
Flatness demands, essentially, that cycles of the system cannot be alternated during an execution.
It is thus a strong restriction but decreases the computational complexity of verification tasks significantly.
To benefit from the improved complexity while being generally applicable, our approach verifies \emph{flat under-approximations} of a specific depth given as parameter.
Similarly to bounded model checking~\cite{DBLP:conf/tacas/BiereCCZ99,DBLP:series/faia/Biere09}, the parameter allows the user to flexibly adjust the trade-off between exhaustiveness and computational effort.
An essential advantage of flat under-approximations is that they represent sets of complete (infinite) runs instead of only a finite number of bounded prefixes.
They can be understood as a bounded unfolding of loop alternations, represented symbolically.
When increasing the approximation depth to include one more alternation, an infinite number of additional runs is represented, and verified at once.
Considering first a small depth and increasing it only if no witness was found allows for finding “simple” witnesses quickly where they exist, even for complex path properties that cannot be evaluated on prefixes.
The underlying theory provides a bound on the maximal depth that needs to be considered in the case of a flat system.
The method is (necessarily) incomplete in the general case but can nevertheless be directly applied.

\subsubsection{Contributions.}
As conceptual basis, we first extend the theory of model-checking counting logics on flat structures developed in~\cite{DBLP:conf/concur/DeckerHLST17}, where only frequency constraints and Kripke structures were considered.
Symbolic models called \emph{augmented path schemas} were introduced to represent sets of runs.
We extend the definitions and techniques to apply to more general counting constraints and flat counter systems while preserving the previous complexity bounds.
This is a consequent continuation of the development of the theory.
From the user perspective, it is a valuable extension, since \myLTL provides a much more flexible specification language and counter systems an extended application domain.
It is particularly important for the practical application of the method.

Subsequently, based on the lifted theory, we describe an explicit formulation of the (approximated) model-checking problem in quan\-ti\-fi\-er-free Presburger arithmetic (\QPA).
Recall that Presburger arithmetic is first-order logic over the integer numbers with addition.
Its satisfiability problem is decidable~\cite{Presburger29} and in the case of the quantifier-free fragment in \NP~\cite{BoroshT76}.
Importantly, the theory of \QPA is well-supported by a number of competitive SMT-solvers (cf.~\cite{DBLP:journals/jar/CokSW15}).
Our construction is parametrised by the depth of the flat approximation that is to be verified, and the resulting \QPA formula is linear in the problem size and the chosen depth.

We have implemented the incremental model-checking procedure based on the \QPA encoding and the z3 SMT solver~\cite{DBLP:conf/tacas/MouraB08}.
Verification tasks of the RERS Challenge~\cite{DBLP:journals/sttt/HowarIMSBP14} and counting variations were used to evaluate the effectiveness of our approach.

\vspace{-3mm}
\subsubsection{Related work.}

In~\cite{DBLP:conf/tase/BolligDL12} an \LTL extension to express relative frequencies, called \fLTL, was studied.
It features a generalised until operator that can be understood as a variant of the $\Uc{.}$ operator restricted to a specific class of counting constraints.
Various other classes were studied in the context of \CTL~\cite{DBLP:journals/corr/abs-1211-4651}.
One of the corresponding \CTL variants, denoted \CCTLpm, admits integer coefficients and thus represents the branching-time analog to \myLTL, although interpreted over finite Kripke structures.
The difference between linear and branching time is crucial, however.
Satisfiability, and hence model checking Kripke structures, is undecidable for \fLTL~\cite{DBLP:conf/tase/BolligDL12} (and hence for \myLTL) but decidable for its branching-time analog \fCTL and even \CCTLpm~\cite{DBLP:journals/corr/abs-1211-4651,DBLP:conf/concur/DeckerHLST17}.
Counting extensions were also studied for regular expressions in~\cite{DBLP:conf/concur/HoenickeMO10,DBLP:conf/fsttcs/AbdullaAMS15}.
The notion of flat (or \emph{weak}) systems was investigated as a sensible restriction to reduce the computational complexity of various verification problems.
Considering (finite) Kripke structures, model-checking \LTL properties, which is \PSpace-complete~\cite{DBLP:journals/jacm/SistlaC85}, becomes \NP-complete under the flatness condition~\cite{DBLP:conf/concur/KuhtzF11}.
It follows from \cite{DBLP:conf/tase/BolligDL12} that model-checking \fLTL, and thus all more expressive counting logics, is undecidable.
Over flat Kripke structures, the problem is in \NExp and even an extremely powerful counting extension of \CTLstar was shown to become decidable~\cite{DBLP:conf/concur/DeckerHLST17}.
A similar impact is observable for (infinite state) counter systems.
While reachability is already undecidable for two-counter systems~\cite{Minsky67}, results from~\cite{DBLP:conf/cav/ComonJ98} provide that flatness recovers decidability with an arbitrary number of counters (see also~\cite{DBLP:conf/csl/ComonC00}).
Later, it was shown in~\cite{DBLP:journals/iandc/DemriDS15} that \LTL properties (including past) can generally be evaluated in \NP (see also \cite{Dhar2014}).
The authors also make the suggestion to consider flat systems as under-approximations, which is addressed here.
Increasing the depth of a flat under-approximation is similar to so-called \emph{loop acceleration} in symbolic verification.
It aims at stepping over an arbitrary number of consecutive iterations of a loop during state space exploration, by symbolically representing its effect.
Since this is particularly effective for simple loops, flatness is a desired property~\cite{DBLP:conf/atva/BardinFLS05} also in this setting.
Unfortunately, acceleration typically concerns the computation of reachability sets~\cite{DBLP:conf/atva/BardinFLS05,DBLP:conf/pldi/BeyerHMR07,DBLP:conf/tacas/CaniartFLZ08,DBLP:journals/fac/KroeningW10,DBLP:conf/atva/HojjatIKKR12}
and is thus insufficient when analysing \emph{path properties} as expressible in (extensions of) \LTL.
For accelerating the latter, flat systems, and path schemas in particular, provide a suitable symbolic model since they represent entire runs.

\subsubsection{Outline.}
\label{ssc:intro-outline}

First, \cref{sec:definitions} provides basic definitions.
In \cref{sec:theory}, a generalised notion of augmented path schemas is introduced and employed to lift the decidability results of~\cite{DBLP:conf/concur/DeckerHLST17}.
It provides the basis for \cref{sec:encoding} describing the parametrised encoding of the model-checking problem into \QPA.
\Cref{sec:implementation} reports on our implementation of the approach and
\cref{sec:conclusion} concludes.
\section{Counting in Linear Temporal Logic}
\label{sec:definitions}

\subsubsection{Constraints and counter systems.}
For $x,y∈ℤ$ let $[x,y]$ denote the (potentially empty) interval $｛x,x+1,…,y｝⊂ℤ$.
A \emph{constraint} over a set $X$ is a linear arithmetic inequation $τ≥b$ where $τ=∑_{i=0}^n a_ix_i$, $n∈ℕ$, $b,a_i∈ℤ$, and $x_i∈X$ for $i∈[0,n]$.
For convenience, we may use relation symbols $≤$, $<$, and $>$, denoting arithmetically equivalent constraints, e.g.\ $2x_1 + x_2<3$ denotes $-2x_1 - x_2≥-2$.
The \emph{dual} of a constraint $τ≥b$ is denoted by $\overline{τ≥b}$ and defined as the equivalent of $τ<b$.
For a valuation $θ: X → ℤ$, we denote by $⟦τ⟧(θ) := ∑_{i=0}^n a_iθ(x_i)$ the arithmetic evaluation of $τ$.
Satisfaction is defined as $θ⊧τ≥b$ if and only if $⟦τ⟧(θ)≥b$.
Constraint sets are interpreted as conjunction and satisfaction is defined accordingly.
The set of all constraints over $X$ is denoted $\Guards(X)$.
For convenience, arithmetic operations are lifted point-wise to integer-valued functions of equal domain.

Let $Λ$ be a set of \emph{labels} and $C_𝓢$ a finite set of \emph{system counters}. A \emph{counter system (CS)} over $Λ$ and $C_𝓢$ is a tuple $𝓢=(S,Δ,s_I,λ)$ where $S$ is a finite set of \emph{control states}, $s_I∈S$ is the \emph{initial state},
$λ:S → 2^Λ$ is a \emph{labelling} function, and
$Δ⊆S×ℤ^{C_𝓢}×2^{\Guards(C_𝓢)}×S$ is a finite set of \emph{transitions} carrying an \emph{update} $μ:C_𝓢 → ℤ$ to the system counters and a finite set of \emph{guards} $Γ⊆\Guards(C_𝓢)$ over them.
A \emph{configuration} of $𝓢$ is a pair $(s,θ)$ comprised of a state $s∈S$ and a \emph{valuation} $θ:C_𝓢 → ℤ$.
A \emph{run} of $𝓢$ is an infinite sequence  $ρ=(s_0,θ_0)(s_1,θ_1)…∈(S×ℤ^{C_𝓢})^ω$ such that $(s_0,θ_0)=(s_I,𝟎)$ and for all positions $i∈ℕ$ there is a transition $(s_i,μ_i, Γ_i, s_{i+1})∈Δ$ such that $θ_{i+1}=θ_i+μ_i$ and $θ_{i+1}⊧Γ_i$.
The set of all runs of $𝓢$ is denoted $\Runs(𝓢)$.

Let $λ^{\#}:S^* → ℕ^{Λ}$ denote the accumulation of labels in a multi-set fashion, counting the number of occurrences of each label on a finite state sequence $w∈S^*$ by $λ^{\#}_𝓟(w): ℓ ↦ |｛i∈[0,|w|-1]｜ℓ∈λ(w(i))｝|$ for all $ℓ∈Λ$.
The set of successors of a state $s∈S$ in $𝓢$ be denoted by $\suc{𝓢}(s):=｛s'∈S ｜∃_{μ,Γ}:(s,μ,Γ,s')∈Δ｝$, and the corresponding transitive and reflexive closure by $\suc{𝓢}^*(s)$.
A (finite) \emph{path} in $𝓢$ is a (finite) state sequence $w=s_0s_1…$
with $s_{i+1}∈\suc{𝓢}(s_i)$ for all $0≤i<|w|$.
A finite path $w=s_0…s_n$ is \emph{simple} if no state occurs twice, it is a \emph{loop} if $s_0∈\suc{𝓢}(s_n)$, and a \emph{row} if no state is part of any loop in $𝓢$.
The counter system $𝓢$ is \emph{flat} if for every state $s∈S$ there is at most one simple loop $s_0…s_n$ with $s_0=s$.
Let the size of $𝓢$ be denoted by $|𝓢|$ and defined as the length of its syntactic representation with numbers encoded binary.

\subsubsection{Counting LTL.}

We consider linear temporal logic extended by counting constraints in the style of~\cite{DBLP:conf/time/LaroussinieMP10}.
In contrast, however, we admit arbitrary \emph{integer coefficients}. Moreover, the semantics is defined in terms of runs of \emph{counter systems}
and the logic provides access to the counter valuation by means of Presburger constraints.
Let $AP$ and $C$ be fixed, finite sets of atomic propositions and counter names, respectively.
The set of \CLTL formulae (denoted simply by $\CLTL$) is defined by the grammar
\begin{align*}
  φ &::= \true ~｜~ p ~｜~ γ ~｜~ φ ∧ φ ~｜~ ¬φ  ~｜~ \X φ ~｜~ φ \Uc{τ≥b} φ \\
  τ &::= a·φ｜τ+τ
\end{align*}
for atomic propositions $p ∈ AP$, guards over counter names $γ∈\Guards(C)$ and integer constants $a,b∈ ℤ$.
Additional abbreviations may be used with expected semantics, in particular $\false:=¬\true$, $φ \U ψ:= φ\Uc{1·\true ≥ 0}ψ$ and $\Fc{τ≥b}φ:=\true\Uc{τ≥b}φ$.
We may write $\myLTL(C')$ for the restriction to formulae that only use counter names from some specific set $C'⊆C$.
By $\sub(φ)$ we denote the set of subformulae of $φ$ (including itself).

Let $𝓢=(S,Δ,s_I,λ)$ be a counter system over counters $C_𝓢$ with a run $ρ=(s_0,θ_0)(s_1,θ_1)…$ and $i≥0$ a position on $ρ$.
Observe that expressions of the form $τ≥b$ are in fact arithmetic constraints from the set $\Guards(\myLTL)$.
The satisfaction relation $⊧$ is defined inductively as follows.
For plain \LTL formulae, the usual definition applies.
Additionally, for $(τ≥b)∈ℭ(\myLTL(C_𝓢))$, $γ∈ℭ(C_𝓢)$, and $φ,ψ∈\myLTL(C_𝓢)$ let
\[\begin{array}{l@{\hspace{1.4ex}}c@{\hspace{1.4ex}}l}
  (𝓢,ρ,i) ⊧ γ     &:⇔& θ_i ⊧ γ \\
  (𝓢,ρ,i) ⊧ φ\Uc{τ≥b}ψ &:⇔& ∃_{j≥i}: (𝓢,ρ,j)⊧ψ \text{ and } ⟦τ⟧(\#^{𝓢,ρ}_{i,j-1})≥b \\
                      && \text{ and } ∀_{i≤k<j}: (𝓢,ρ,k) ⊧ φ
\end{array}\]
where $\#_{i,j}^{𝓢,ρ}:\myLTL → ℕ$ denotes the function mapping a \myLTL formula $χ$ to the number
\[
  \#_{i,j}^{𝓢,ρ}(χ):=|｛k｜i≤k≤j, (𝓢,ρ,k)⊧χ｝|
\]
of positions on $ρ$ between $i$ and $j$ satisfying it.
Notice that this is well-defined because the mutual recursion descends towards strict subformulae.
We write $(𝓢,ρ)⊧χ$ if $(𝓢,ρ,0)⊧χ$ and $𝓢⊧χ$ if there is $ρ∈\Runs(𝓢)$ with $(𝓢,ρ)⊧χ$.

The logic \fLTL~\cite{DBLP:conf/tase/BolligDL12} features a dedicated \emph{frequency-until} operator $\U^{\frac{a}{b}}$ for $a,b∈ℕ$ and $a≤b>0$ that can be considered as restricted variant of $\Uc{.}$.
An \fLTL formula $φ\U^{\frac{a}{b}}ψ$ specifies that a formula $φ$ holds at least at a fraction $0≤\frac{a}{b}≤1$ of all positions before some position satisfying $ψ$.
This is equivalently expressed in \cLTL by $\true \Uc{b·φ - a·\true≥0}ψ$.

\vspace{-3mm}
\subsubsection{Model checking.}

We target the \emph{existential model-checking problem} for \CLTL.\@
Given a counter system $𝓢$ and a \CLTL formula $Φ$ the task is to decide whether $𝓢⊧Φ$, i.e., to compute if $𝓢$ contains a run satisfying $Φ$.
The problem is undecidable for two reasons:
First, counter systems extend Minsky machines~\cite{Minsky67} and even \LTL can express their undecidable (control-state) reachability problem.
Second, \CLTL extends \fLTL and checking a universal Kripke structure encodes its undecidable satisfiability problem~\cite{DBLP:conf/tase/BolligDL12}.
We therefore approach a parametrised approximation of the problem that considers only runs with a specific shape, namely those represented by
so-called \emph{path schemas}.
A path schema~\cite{DBLP:conf/concur/LerouxS04,DBLP:journals/iandc/DemriDS15} is characterised by a (connected) sequence $u_0v_0u_1v_1…u_nv_n$ of paths $u_i$ and cycles $v_i$ of $𝓢$.
It represents all those runs $ρ$ of $𝓢$ that traverse a state sequence of the form $u_0v_0^{ℓ_0}…u_{n-1}v_{n-1}^{ℓ_{n-1}}u_nv_n^ω$.
Restricting the length of such a schema effectively controls how complicated the shape of the considered runs can be.
In particular, it bounds the cycle alternation performed by a run.

\begin{definition}[Flat model checking]
  Let $𝓢=(S,Δ,s_I,λ)$ be a counter system and $n∈ℕ$.
  The \emph{flat approximation} of depth $n$ of $𝓢$ is the set  $\FA(𝓢,n)⊆\runs(𝓢)$ such that, for all $ρ=(s_0,θ_0)(s_1,θ_1)…∈\Runs(𝓢)$,
\begin{align*}
  ρ∈\FA(𝓢,n) ~⇔~
    & ∃_{u_0,v_0,…,u_m,v_m∈S^*}: |u_0v_0u_1v_1…u_mv_m|≤n \\
    & ~ ∧ ∃_{k_0,…,k_{m-1}∈ℕ}: s_0s_1… = u_0v_0^{k_0}\, … \, u_{m-1}v_{m-1}^{k_{m-1}}\, u_mv_m^ω.
\end{align*}
  The \emph{flat model-checking problem} is to decide for a given \cLTL formula $φ$, whether there is a run $ρ∈\FA(𝓢,n)$ with $(𝓢,ρ)⊧φ$, denoted $\FA(𝓢,n)⊧φ$.
\end{definition}

A flat approximation $\FA(𝓢,n)$ induces a flat counter system $𝓕$ such that $\FA(𝓢,n)= \runs(𝓕)$ and thus a series $(𝓕_n)_{n∈ℕ}$ of flat counter systems representing an increasing number of runs of $𝓢$.
Flat model checking can hence be understood as verifying the $n$th system in this series providing the computational benefits of flatness in the concrete case.
As mentioned earlier, this is similar to bounded model checking, where the approximation is prefix-based and represents only a finite number of runs.

\section{Model Checking \CLTL over Flat Counter Systems}
\label{sec:theory}

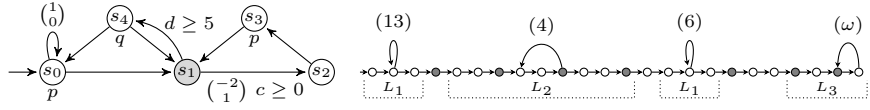
\begin{figure}[tb]

\begin{tikzpicture}[
    baseline = (s0),
    small nodes,
    node distance = 5.5em,
            every edge/.append style={inner sep=2pt}
]

\tikzset{
  s0/.style = {"{$p$}" {below, inner sep = 1pt}},
  s1/.style = {fill={black!15}},
  s2/.style = {},
  s3/.style = {"{$p$}" {below, inner sep = 1pt}},
  s4/.style = {"{$q$}" {below, inner sep = 1pt}}
}

\node[state, initial, s0] (s0) {$s_0$};
\node[state, right of = s0, s1] (s1) {$s_1$};
\node[state, right of = s1, s2] (s2) {$s_2$};

\path (s1) -- node[yshift = 2.2em, state, s3] (s3) {$s_3$} (s2);

\node[state, left of = s3, s4] (s4) {$s_4$};

\draw[->] (s0) edge[loop above,  "{$\binom{1}{0}$}"] (s0)
          (s0) edge (s1)
                    (s1) edge ["{$\binom{-2}{1}$}~{$c≥0$}"']  (s2)
          (s1) edge [bend right, "{$d≥5$}"', inner sep = 0pt] (s4)
          (s2) edge (s3)
          (s3) edge (s1)
          (s4) edge (s1)
          (s4) edge (s0);

\end{tikzpicture}
\begin{tikzpicture}[micro nodes, baseline = (q0)]
\node[state, initial] (q0) {};
\node[state, right of = q0] (q1) {};
\node[state, right of = q1] (q2) {};
\node[state, right of = q2, fill=gray] (q3) {};
\node[state, right of = q3] (q4) {};
\node[state, right of = q4] (q5) {};
\node[state, right of = q5, fill=gray] (q6) {};
\node[state, right of = q6] (q7) {};
\node[state, right of = q7] (q8) {};
\node[state, right of = q8, fill=gray] (q9) {};
\node[state, right of = q9] (q10) {};
\node[state, right of = q10] (q11) {};
\node[state, right of = q11, fill=gray] (q12) {};
\node[state, right of = q12] (q13) {};
\node[state, right of = q13] (q14) {};
\node[state, right of = q14] (q15) {};
\node[state, right of = q15] (q16) {};
\node[state, right of = q16, fill=gray] (q17) {};
\node[state, right of = q17] (q18) {};
\node[state, right of = q18] (q19) {};
\node[state, right of = q19, fill=gray] (q20) {};
\node[state, right of = q20] (q21) {};
\node[state, right of = q21, fill=gray] (q22) {};
\node[state, right of = q22] (q23) {};

\path[->] (q0) edge (q1)
          (q1) edge (q2)
          (q2) edge (q3)
          (q3) edge (q4)
          (q4) edge (q5)
          (q5) edge (q6)
          (q6) edge (q7)
          (q7) edge (q8)
          (q8) edge (q9)
          (q9) edge (q10)
          (q10) edge (q11)
          (q11) edge (q12)
          (q12) edge (q13)
          (q13) edge (q14)
          (q14) edge (q15)
          (q15) edge (q16)
          (q16) edge (q17)
          (q17) edge (q18)
          (q18) edge (q19)
          (q19) edge (q20)
          (q20) edge (q21)
          (q21) edge (q22)
          (q22) edge (q23)
          (q1) edge [loop above left] node[swap, auto, font=\scriptsize] {$(13)$} (q1)
          (q9) edge [bend right=80, looseness = 2] node[auto, swap, font=\scriptsize] {$(4)$} (q7)
          (q15) edge [loop above left] node[swap, auto, font=\scriptsize] {$(6)$} (q15)
          (q23) edge[bend right=90, looseness = 4] node[auto, swap, font=\scriptsize] {$(ω)$} (q22);

\draw[densely dotted] ([xshift=-0.5ex,yshift=-0.5ex] q0.south west) |- ([xshift=0.5ex,yshift=-2.2ex] q2.south east) -- ++(0,1.7ex);
\draw[densely dotted] ([xshift=-0.5ex,yshift=-0.5ex] q4.south west) |- ([xshift=0.5ex,yshift=-2.2ex] q12.south east) -- ++(0,1.7ex);
\draw[densely dotted] ([xshift=-0.5ex,yshift=-0.5ex] q14.south west) |- ([xshift=0.5ex,yshift=-2.2ex] q16.south east) -- ++(0,1.7ex);
\draw[densely dotted] ([xshift=-0.5ex,yshift=-0.5ex] q20.south west) |- ([xshift=0.5ex,yshift=-2.2ex] q23.south east) -- ++(0,1.7ex);

\node[anchor = north, inner sep = 2pt] at (q1.south) {$L_1$};
\node[anchor = north, inner sep = 2pt] at (q8.south) {$L_2$};
\node[anchor = north, inner sep = 2pt] at (q15.south) {$L_1$};
\path (q21.south) -- node[anchor = north, inner sep = 2pt] {$L_3$} (q22.south);

\end{tikzpicture}
   \caption{A counter system $𝓢$ over propositions $AP=｛p,q｝$ as labels and counters $｛c,d｝$, and (a sketch of) an APS $𝓟$ in $𝓢$ that alternates the loops $L_1=s_0$ and $L_2=s_1s_2s_3$ of $𝓢$.
  Associating with each loop of $𝓟$ a number of iterations (potentially) identifies one specific run of $𝓢$ that is represented by $𝓟$.\label{fig:example-counter-system}}
  \vspace{-5mm}
\end{figure}

This section is dedicated to lifting the technique for model-checking \fLTL over flat Kripke structures~\cite{DBLP:conf/concur/DeckerHLST17} to \CLTL and flat counter systems.
The central aspect is the definition of \emph{augmented path schemas (APS)} and the notion of \emph{consistency}.
We observe that consistent APS are suitable witnesses for runs because they are of bounded size and exist if a formula is satisfied.
The \QPA encoding of the flat-model-checking problem presented in \cref{sec:encoding} builds on these definitions.
To simplify notation, we fix in this section a counter system $𝓢=(S_𝓢,Δ_𝓢,s_I,λ)$ and a \CLTL formula $Φ$, both over counters $C_𝓢$.
Augmented path schemas~\cite{DBLP:conf/concur/DeckerHLST17} extend path schemas by a labelling that provides additional information, as well as counters and guards to constrain the set of runs of an APS beyond a specific shape.
The following definition extends that of~\cite{DBLP:conf/concur/DeckerHLST17} to take the counters and guards of $𝓢$ into account.
See \cref{fig:example-counter-system} for an example.

\begin{definition}[APS\label{def:aps}]
  An \emph{augmented path schema (APS)} in $𝓢$ is a structure $𝓟=(Q, Δ_𝓟, λ_𝓟, \org)$ where
\begin{itemize}
    \item $(Q, Δ_𝓟,q_0, λ_𝓟)$ is a flat counter system over $Q=｛q_0,…,q_n｝$, for some $n∈ℕ$, with labelling $λ_𝓟: Q → 2^{\sub(Φ)∪AP}$ and simple path $q_0…q_n$;
    \item $\org: Q → S_𝓢$ maps every state to an \emph{origin} such that $λ_𝓟(q)∩AP=λ_𝓢(\org(q))∩AP$ and $\org(q_0)=s_I$;
    \item for each transition $(q,μ,Γ,q')∈Δ_𝓟$ there is       $(\org(q),\hat{μ},\hat{Γ},\org(q'))∈Δ_𝓢$ with $\hat{Γ}⊆Γ$ and $\hat{μ}(c)=μ(c)$ for all $c∈C_𝓢$;
          \item $Δ_𝓟=Δ_{\text{fwd}}\operatorname{\dot{∪}}Δ_{\text{bwd}}$ is comprised of \emph{forward-} and \emph{backward transitions} where
                \begin{itemize}
  \item $Δ_{\text{fwd}}=｛(q_0,μ_0,Γ_0,q_{1}),…,(q_{n-1},μ_{n-1},Γ_{n-1},q_{n})｝$,
  \item there is $(q_n,μ_n,Γ_n,q_{n'})∈Δ_{\text{bwd}}$, for $n'≤n$, closing the \emph{last loop}, and
  \item         for all $(q_j,μ,Γ,q_i),(q_k,μ',Γ',q_h)∈Δ_{\text{bwd}}$ we have $i≤j$, $h≤k$, and
                the corresponding loops $q_hq_{h+1}…q_k$ and $q_iq_{i+1}…q_j$ are disjoint; and
\end{itemize}
    \item for each loop $L=q_iq_{i+1}…q_{i+ℓ}$ there is a \emph{front row} $F=q_{i-ℓ-1}…q_{i-1}$ and, if $i+ℓ<n$, a \emph{rear row} $R=q_{i+ℓ+1}…q_{i+2ℓ+1}$ with identical labelling
    $λ_𝓟(q_{i-ℓ-1})…λ_𝓟(q_{i-1})=λ_𝓟(q_i)…λ_𝓟(q_{i+ℓ}) = λ_𝓟(q_{i+ℓ+1})…λ_𝓟(q_{i+2ℓ+1})$.
  \end{itemize}
\end{definition}

The paths, loops, rows, and runs of $𝓟$ are those of the underlying counter system where the latter are restricted to those visiting the last state $q_n$ of $𝓟$.
The mapping $\org$ is lifted from states to paths and runs as expected, restricting the valuations to the counters $C_𝓢$ of $𝓢$.
Then, for every run $ρ$ of $𝓟$, the sequence $\org(ρ)$ is a run of $𝓢$ starting in $\org(q_I)=s_I$.
We denote by $\lastloop(𝓟) := q_{n'}…q_n$ the last loop of $𝓟$.
Observe that the definition requires each loop to be preceded and (except for $\lastloop(𝓟)$) succeeded by state sequences that may be considered as an \emph{unfolding} regarding the labelling sequence.
These front and rear rows are needed for technical reasons to cover edge-cases in reasoning on the first and last loop iteration, respectively.

We are interested in APS that provide a semantically correct labelling because they allow us to reason syntactically on where a particular formula is satisfied.

\begin{definition}[Correctness]
    A state $q∈Q$ of an APS $𝓟$ is \emph{correctly labelled} with respect to a \CLTL formula $φ∈\sub(Φ)$ if for all runs
  $ρ=(q_0,θ_0)(q_1,θ_1)…∈\Runs(𝓟)$ and all positions $x∈ℕ$ with $q_x=q$ we have $(𝓢,\org(ρ),x)⊧φ ⇔ φ∈λ_𝓟(q)$.
\end{definition}
This notion is very strict in the sense that the annotation must \emph{always} be in line with the \CLTL semantics.
Observe that there may not even exist a correct labelling for a particular state: if the latter resides on a loop it may occur more than once on some run and a formula $Φ$ may hold at one of them but not at the other (e.g., because $Φ$ imposes a minimal number of iterations to follow).
However, an APS in $𝓢$ that is actually correctly labelled witnesses the existence of a run satisfying $Φ$ in case it is non-empty and its initial state is labelled by $Φ$.
In~\cite{DBLP:conf/concur/DeckerHLST17}, the syntactic criterion called \emph{consistency} was introduced in order to characterise APS
that are labelled correctly with respect to \fLTL formulae.
We generalise the definition and the results to \CLTL, i.e., from relative frequencies to arbitrary linear constraints and from Kripke structures to counter systems.

Consider an APS $𝓟=(Q,Δ_𝓟,λ_𝓟, \org)$ using counters $C_𝓟⊇C_𝓢$ where $q_0…q_n$ is the unique simple path traversing all states of $𝓟$.
The criterion distinguishes the syntactical forms of a \CLTL formula based on the top most operator and identifies for each case syntactical conditions that certify satisfaction or violation of a corresponding formula. Further subordinate cases formulate individual conditions to matching the various situations that may apply to a control state, e.g., whether it is on a loop or not.
Before presenting the formal definition, let us discuss the rationale of the individual conditions.

\vspace{-3mm}
\subsubsection{Consistency for non-until formulae.}

The simplest case is that of propositions, because these labels are correct by definition.
Recall that constraints $γ∈\Guards(C_𝓢)$ over system counters, e.g.\ $c_1 -2 c_2 ≥ 0$, are not only valid atomic \CLTL formulae but also valid transition guards.
Therefore, the reasoning on their satisfaction can directly be moved to the level of the counter system.
If all incoming transitions of a state $q∈Q$ are guarded by some constraint $γ$, then every valid run necessarily satisfies it whenever visiting $q$.
Similarly, if these transitions are guarded by the dual constraint $\overline{γ}$, then $γ$ can not hold at any occurrence of $q$ on any run.

If $Φ$ is a Boolean combination, correctness can be established locally for any state $q$ when inductively assuming that $q$ is labelled correctly by all the strict subformulae.
For example, a negation $¬φ$ holds on all runs at all positions of a state $q$ if and only if on all runs $φ$ does not hold at $q$.
With the assumption that the labelling with respect to $φ$ is correct, labelling $q$ by $¬φ$ is correct if and only if $q$ is not labelled by $φ$, and vice versa.
Similar reasoning applies to conjunctions and the temporal operator $\X$.

\vspace{-3mm}
\subsubsection{Consistency for until formulae using balance counters.}
For counted until formulae, we also make use of the counting capabilities of the system model, although the reasoning is more involved.
Consider $Φ$ to have the form $φ\Uc{τ≥b}ψ$, and let $q,q'∈Q$ be row states such that $q'∈\suc{𝓟}^*(q)$ is a successor of $q$ and (correctly) labelled by $ψ$.
Assume that the states in-between $q$ and $q'$ are correctly labelled by $φ$.
In order to establish that $Φ$ holds at state $q$ on any run, it remains to enforce the counting constraint on the intermediate segment.
To this end, also assume that $𝓟$ features a counter $c_{τ,q}$ that tracks the value of the term $τ$ as a \emph{balance} that starts with zero at $q$ and is updated according to the effect that each individual state would have on the value of $τ$.
For example, if $τ=p_1-2p_2$, then the counter is updated by $+1$ on every outgoing transition of a state labelled by $p_1$, because this is what each such state contributes to the term value.
The counter would be update by $-2$ on the outgoing transitions, if the state is labelled by $p_2$, and consequently by $1-2=-1$ if it carries both labels.
Then, upon reaching $q'$ along some run, the counter $c_{τ,q}$ would hold precisely the value of the counting term $τ$ evaluated on the intermediate path taken from $q$ to $q'$.
If the incoming (forward) transition of $q'$ is now labelled by the guard $c_{t,q}≥b$, then $Φ$ can be assumed to hold whenever a valid run visits $q$ because $q'$ is certainly visited and will then serve as witness.
Dually, if \emph{all} such potential witness states $q'$ are guarded instead by the dual constraint $c_{t,q}<b$, then there is no way a valid run could satisfy $Φ$ when visiting $q$.

\begin{definition}[Balance counter\label{def:balance-counter}]
  Let $𝓟=(Q,Δ_𝓟,λ_𝓟, \org)$ be an APS in $𝓢$ with counters $C_𝓟$.
  Let $τ$ be a constraint term over $\sub(Φ)$, and $q∈Q$ a row state in $𝓟$.
  A \emph{balance counter} for $τ$ and $q$ in $𝓟$ is a counter $c_{τ,q}∈C_𝓟$ that is updated, on all transitions $(q_1,μ,Γ,q_2)∈Δ_𝓟$, by
\[
  μ(c_{τ,q}) = \begin{cases}
    0 & \text{if $q_1∉\suc{𝓟}^*(q)$} \\
    ⟦τ⟧(λ^{\#}_𝓟(q_1)) & \text{otherwise}.
  \end{cases}
\]
\end{definition}

In combination with appropriately guarded states, balance counters allow us to reason syntactically about the satisfaction of $Φ$.
Such counters are particularly useful to track the value of a term across an entire loop, even if some runs of $𝓟$ iterate it more often than others.

\vspace{-3mm}
\subsubsection{Static consistency conditions for until formulae.}
If there is no entire loop between two states $q$ and $q'$, using a counter is still possible but not necessary.
Each run passes precisely once the (unique) path between $q$ and $q'$, so whether or not $q'$ witnesses satisfaction of $Φ$ at $q$ can be determined statically, independently of the precise course of the run in other parts.
While the existence of a balance counter and appropriate guards imply that a formula is satisfied, it would be too restrictive to consider this as only option.
There are situations where satisfaction of a formula can not be witnessed by a balance counter.
For example, if a witness state $q'$ is part of a loop, a corresponding guard may be satisfied at one of its occurrences on a run but not at all of them.
While the consistency criterion is intended to be strong enough to imply correctness, it shall also admit a sufficiently large class of APS to represent all reasons for satisfaction (and violation).
Therefore, the definition admits also the static reasoning.

A further case treated explicitly concerns the effect of the \emph{last loop}.
If traversing it once exhibits a positive effect on the evaluation of $τ$, then it dominates the effects of all other loops, since it is traversed infinitely often.
Therefore, if it can be reached from $q$ and traversed once without violating $φ$, and contains some witness state labelled by $ψ$, then $Φ$ is necessarily satisfied when a run reaches $q$.

Finally, the last case considered by the consistency criterion is concerned with the satisfaction of $Φ$ when visiting states that are situated directly on a loops: If $Φ$ holds at the first occurrence of a state $q$ on a run \emph{and} at the last, then the formula holds also at all occurrences of $q$ in-between.
The reason is, essentially, that the effect of one iteration of a loop on the value of the term $τ$ is always the same (at least, if the labelling by all subformulae is correct, as we have assumed).
Therefore, the worst (i.e., smallest) value of $τ$
is encountered either in the first or the last iteration.
Augmented path schemas are defined to feature for each loop a preceding and a succeeding row that are exact copies and can be considered as \emph{unfoldings}.
Hence, if these are correctly labelled with respect to $Φ$, then the loop labelling inherits their correctness.

Using the above reasoning, it can be shown that the following definition of consistency is a sufficient criterion for correctness.
It extends that of \cite{DBLP:conf/concur/DeckerHLST17} to the present context and accounts for the various subtleties arising from the different cases.

\begin{definition}[Consistency]
 \label{def:consistency}
Let $𝓟=(Q,Δ_𝓟,λ_𝓟, \org)$ be an APS in $𝓢$ with $|Q|=n$, simple path $q_0…q_{n-1}$, and $φ$ a \cLTL formula.
A state $q_i∈Q$ is \emph{$φ$-consistent} if $φ∈AP$ is an atomic proposition or
    \begin{enumerate}[(A)]
                          \item\label[condition]{itm:consistency-constraint}
        $φ=(τ≧b)∈\Guards(C)$,         all incoming transitions $(q,μ,Γ,q_i)∈Δ_𝓟$ are guarded by $φ∈Γ$ if $φ∈λ_𝓟(q_i)$ and by $\overline{φ}∈Γ$ otherwise, and if $i=0$, then $φ∈λ_𝓟(q_i) ⇔ 0≥b$.
    \setcounter{resumeenum}{\value{enumi}}
  \end{enumerate}
For non-atomic formulae $φ$, the state $q_i$ is \emph{$φ$-consistent} if for all $ψ∈\sub(φ)∖｛φ｝$ all states $q∈Q$ are $ψ$-consistent and one of the following \cref{itm:consistency-bool,itm:consistency-X,itm:consistency-U} applies.
\begin{enumerate}[(A)]
  \setcounter{enumi}{\value{resumeenum}}
  \item\label[condition]{itm:consistency-bool}
    $φ=χ∧ψ$ and $φ∈λ_𝓟(q_i) ⇔ χ,ψ∈λ_𝓟(q_i)$; or
    $φ=¬ψ$ and $¬ψ∈λ_𝓟(q_i) ⇔ ψ∉λ_𝓟(q_i)$.
  \item\label[condition]{itm:consistency-X}
   $φ=\X ψ$ and $\Xψ∈λ_𝓟(q_i) ⇔ ψ∈λ_𝓟(q)$, for all $q∈\suc{𝓟}(q_i)$.

   \item\label[condition]{itm:consistency-U}
     $φ=χ \Uc{τ≧b} ψ$ and one of the following holds:
     \begin{enumerate}[1.]
       \item\label[condition]{itm:consistency-U-goodlast}
         $φ∈λ_𝓟(q_i)$,
         $⟦τ⟧(λ_𝓟^\#(\lastloop(𝓟))>0$,          $ψ∈λ_𝓟(q)$ for some $q∈\lastloop(𝓟)$, and $χ∈λ_𝓟(q')$ for all $q'∈\suc{𝓟}^*(q_i)$.
       \item\label[condition]{itm:consistency-U-counter}
        The state $q_i$ is not part of a loop.
        If $φ∉λ_𝓟(q_i)$, then $ψ∉λ_𝓟(q_i)$ or $0<b$. Further, if $φ∉λ_𝓟(q_i)$, then
            \begin{enumerate}[(i)]
              \item \label[condition]{itm:consistency-U-counter-stat}
                  there is some $k≥i$ such that $χ∉λ_𝓟(q_k)$ and,
                  for each $j∈[i,k]$, $|\suc{𝓟}(q_j)|=1$ and $ψ∈λ_𝓟(q_j) ⇒ ⟦τ⟧(λ_𝓟^\#(q_i…q_{j-1}))<b$ or
              \item \label[condition]{itm:consistency-U-counter-dyn}
                  $𝓟$ contains a balance counter $c_{τ,i}∈C_𝓟$ for $τ$ and $q_i$, and the guard $(c_{τ,i} ≪ b) ∈ Γ$ for all $(q,μ,Γ,q_j)∈Δ_𝓟$ where $j>i$, $ψ∈λ_𝓟(q_j)$, and $∀_{k∈[i,j-1]}:χ∈λ_𝓟(q_k)$.
              \setcounter{resumeenum}{\value{enumiii}}
            \end{enumerate}
          If $φ∈λ_𝓟(q_i)$, then there is $k≥i$ with $ψ∈λ_𝓟(q_k)$, $∀_{j∈[i,k-1]}: χ∈λ_𝓟(q_j)$, and
            \begin{enumerate}[(i)]
              \setcounter{enumiii}{\value{resumeenum}}
              \item \label[condition]{itm:consistency-U-counter-wit-stat}
                $⟦τ⟧(λ_𝓟^\#(q_i…q_{k-1}))≥b$ and $∀_{j∈[i,k-1]}: |\suc{𝓟}(q_j)|=1$, or
              \item \label[condition]{itm:consistency-U-counter-wit-dyn}
                $k>i$ and $𝓟$ contains a balance counter $c_{τ,i}∈C_𝓟$ for $τ$ and $q_i$, and
                the unique transition from $q_{k-1}$ to $q_{k}$ has the form $(q_{k-1},μ,Γ∪｛c_{τ,i}≧b｝, q_k)∈Δ_𝓟$.
                            \end{enumerate}
       \item\label[condition]{itm:consistency-U-unfolding} $q_i$ is on some loop $L$ of $𝓟$, and $q_{i-|L|}$ and $q_{i+|L|}$ (if $L≠\lastloop(𝓟)$) are $φ$-consistent.
       \end{enumerate}
\end{enumerate}
The APS $𝓟$, a loop, or a row in $𝓟$ are $φ$-consistent if all their states are $φ$-consistent, respectively.
\end{definition}

Using a structural induction on a \cLTL formula $φ$ we can show that if some state of an APS is $φ$-consistent, then the state is correctly labelled by that formula.
The base cases those of atomic propositions and guards, concerning \cref{itm:consistency-constraint} of \cref{def:consistency}.
The remaining conditions cover the inductive cases for the potential shape of $φ$ and rely on the fact that the definition demands all states to be consistent with respect to each strict subformula of $φ$.
The proof relies on a thorough investigation of each syntactic case in combination with various specific situations that states can be found in, as discussed above.
It has to deal with the sometimes quite subtle interplay between temporal counting constraints and iterated loops and we omit the technicalities of the proof here in favour of conciseness.

\begin{theorem}[Correctness\label{thm:correctness}]
  If a state $q$ of an APS $𝓟$ in $𝓢$ is $φ$-consistent, then it is labelled correctly with respect to $φ$.
\end{theorem}
Consequently, a non-empty APS in $𝓢$ of which the initial state is $Φ$-consistent and labelled by $Φ$ witnesses that $𝓢⊧Φ$.

\subsubsection{Existence of consistent APS in flat systems.}
Although consistency imposes a very specific shape, it can be shown that for a significant class of systems there is always a $Φ$-consistent APS (of bounded size) if the formula $Φ$ is satisified.
The construction for \fLTL over flat Kripke structures~\cite{DBLP:conf/concur/DeckerHLST17} extends with \cref{def:consistency} to \CLTL.\@

Assume $𝓢$ is flat and let $σ∈\runs(𝓢)$ be a run that satisfies $Φ$.
In the following we sketch how to construct a $Φ$-consistent APS in $𝓢$ that contains (a representation of) $σ$ and is thus labelled by $Φ$ at its initial state.
It is known that each path in a flat structure can be represented by some path schema of linear size \cite{DBLP:conf/atva/BardinFLS05,DBLP:journals/iandc/DemriDS15}.
Hence, let $𝓟$ be an APS containing a run $ρ∈\runs(𝓟)$ with $\org_𝓟(ρ)=σ$ and thus satisfying $Φ$.
The states of this APS can now recursively be labelled by the subformulae of $Φ$ as semantically determined by $ρ$.

The conditions of \cref{def:consistency} can be realised for $Φ$ under the assumption, that the labelling has been completed for each strict subformula.
The construction distinguishes which case applies to $Φ$.
If $Φ$ is an atomic proposition, nothing needs to be done since the labelling is consistent by definition.
Boolean combinations can be realised by simply adjusting the labelling locally for each state of $𝓟$, e.g., including $Φ=¬φ$ in the labelling of a state if and only if it is not labelled by $φ$.
Assume $Φ$ has the form $\Xφ$.
Depending on whether the successor states of a state $q$ are labelled by $φ$ or not, $q$ is labelled by $\Xφ$ or not.
Notice that all successors of a state have the same labelling because either there is only one or the state is the last state of some loop.
In the latter case, the successors are the first states of the loop and its rear copy and thus share the same labelling (cf.~\cref{def:aps}).

For the remaining types of formulae, i.e., until formulae and constraints over system counters, the structure of $𝓟$ may have to be altered, in order to provide a consistent labelling \emph{and} to retain a valid run $ρ$ (as representation of $σ$).
The essential difficulties concern loop states because these may occur at more than one position on $ρ$.
A subformula $φ$ may then be satisfied at some, but not all of these positions.
For example, consistency for a constraint formula $γ=τ≥b$ and a state $q$ demands to add $γ$ or its dual to every incoming transition of $q$, depending on whether we want to label it by $γ$ or not.
Clearly, the guards can simply be added and this would settle consistency.
However, if $γ$ is satisfied at one occurrence of $q$ on $ρ$ but not at another, the guards would be violated at one of these positions and $ρ$ would not be valid anymore.
To establish consistency for until formulae, we may have to add a fresh balance counter to the system and similar issues may arise.
It may therefore be necessary to introduce copies of a state in order to distinguish the positions of the state and label them differently in the APS.\@
The important observation is that during the iteration of a loop the validity of a formula $φ$ at some state switches at most once, assuming the APS is labelled consistently by all subformulae already.
Therefore, loops may have to be \emph{duplicated} once for each subformula, one copy where on all iterations $φ$ holds and one where it does not.
The recursive labelling procedure may therefore increase the size of $𝓟$ exponentially.

\begin{theorem}[Existence\label{thm:existence}]
  If $𝓢$ is flat and $𝓢⊧Φ$ then there is a non-empty and $Φ$-consistent APS in $𝓢$ with initial state labelled by $Φ$ and of at most exponential size in $𝓢$ and $Φ$.
\end{theorem}
Notice that, even if $𝓢$ is \emph{not flat}, each run contained in the flat approximation $\FA(𝓢,n)$ of $𝓢$ can by definition be represented by an APS in $𝓢$ of size $n$.
Therefore, the construction applied to $\FA(𝓢, n)$ also yields an exponential witness.
\begin{corollary}\label{cor:existence}
    If $\FA(𝓢,n)⊧Φ$ then there is a non-empty and $Φ$-consistent APS in $𝓢$ with initial state labelled by $Φ$ and of at most exponential size in $n$ and $Φ$.
\end{corollary}

\section{From Flat Model Checking to Presburger Arithmetic}
\label{sec:encoding}

For solving the flat model-checking problem of a counter system $𝓢=(S,Δ,s_I,λ)$ over counters $C_𝓢$ and a $\CLTL(C_𝓢)$ formula $Φ$, the developments in the previous section devise the search for an augmented path schema $𝓟$ in $𝓢$ that is $Φ$-consistent, labelled initially by $Φ$ and non-empty.
In the following we sketch \footnote{Details are presented in \cref{apx:encoding-details}.}
 a formulation of this search in quantifier-free Presburger arithmetic, aiming at an SMT-based implementation.

The idea is to encode an APS of size $n∈ℕ$ and a run of it as valuation of a set of first-order variables.
We construct a formula $\mathrm{fmc}(𝓢,Φ,n)$ that is satisfiable if there is a run $ρ∈\FA(𝓢,n)$ satisfying $Φ$ and such that any solution represents a valid witness that $𝓢⊧Φ$.
Without restriction, we need only to represent APS $𝓟=(Q,Δ_𝓟,λ_𝓟,\org)$ where the states are natural numbers $Q=[0,n-1]$.
The natural ordering implicitly determines the unique maximal simple path in $𝓟$.
It hence suffices to encode explicitly the beginning and end of loops, the origin and labelling of each state, as well as a valid run.
Further, the formula expresses the satisfaction of all encountered guards and the consistency criterion.

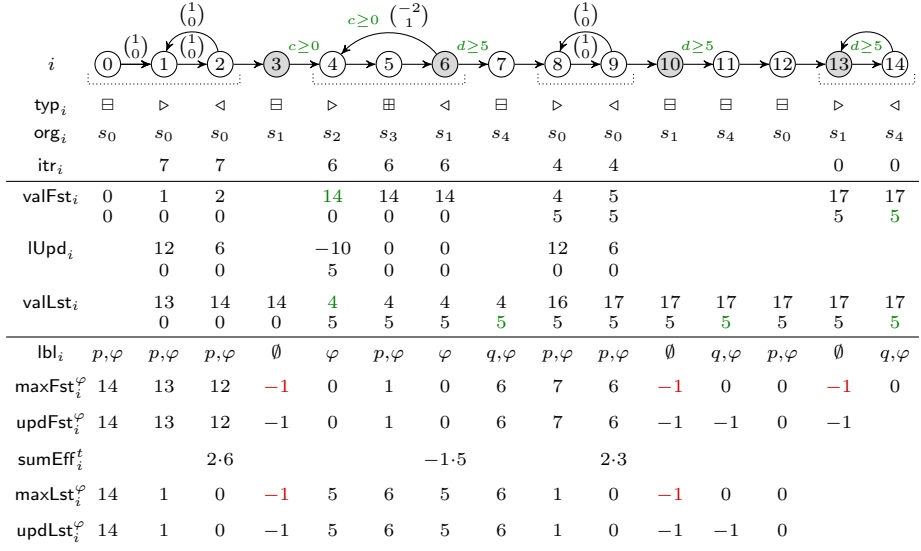
\begin{figure}[tb]

\begin{tikzpicture}

\tikzset{
  s0/.style = {},
  s1/.style = {fill={black!15}},
  s2/.style = {},
  s3/.style = {},
  s4/.style = {},
  sat/.style={color={mygreen}},
  unsat/.style={color={myred}},
}

\matrix [
  font=\scriptsize,
  matrix of math nodes,
  row 1/.style={every node/.style = {state}, small nodes},
  row 1 column 1/.style ={every node/.style={}},
  column sep={2.3em,between origins},
    row sep={1.5ex},
    nodes in empty cells,
    nodes ={inner sep=0pt},
  ] (m)
  {	$i$	&	|[s0]|	0	&	|[s0]|	1	&	|[s0]|	2	&	|[s1]|	3	&	|[s2]|	4	&	|[s3]|	5	&	|[s1]|	6	&	|[s4]|	7	&	|[s0]|	8	&	|[s0]|	9	&	|[s1]|	10	&	|[s4]|	11	&	|[s0]|	12	&	|[s1]|	13	&	|[s4]|	14	\\	[2pt]
	\sym{typ}_i	&		⊟	&		▹	&		◃	&		⊟	&		▹	&		⊞	&		◃	&		⊟	&		▹	&		◃	&		⊟	&		⊟	&		⊟	&		▹	&		◃	\\
	\sym{org}_i	&		s_0	&		s_0	&		s_0	&		s_1	&		s_2	&		s_3	&		s_1	&		s_4	&		s_0	&		s_0	&		s_1	&		s_4	&		s_0	&		s_1	&		s_4	\\
	\sym{itr}_i	&			&		7	&		7	&			&		6	&		6	&		6	&			&		4	&		4	&			&			&			&		0	&		0	\\
		\sym{valFst}_i	&		0	&		1	&		2	&			&	|[sat]|	14	&		14	&		14	&			&		4	&		5	&			&			&			&		17	&		17	\\	[-1ex]
		&		0	&		0	&		0	&			&		0	&		0	&		0	&			&		5	&		5	&			&			&			&		5	&	|[sat]|	5	\\	[1pt]
		\sym{lUpd}_i &			&		12	&		6	&			&		-10	&		0	&		0	&			&		12	&		6	&			&			&			&			&			\\	[-1ex]
		&			&		0	&		0	&			&		5	&		0	&		0	&			&		0	&		0	&			&			&			&			&			\\	[1pt]
		\sym{valLst}_i	&			&		13	&		14	&		14	&	|[sat]|	4	&		4	&		4	&		4	&		16	&		17	&		17	&		17	&		17	&		17	&		17	\\	[-1ex]
		&			&		0	&		0	&		0	&		5	&		5	&		5	&	|[sat]|	5	&		5	&		5	&		5	&	|[sat]|	5	&		5	&		5	&	|[sat]|	5	\\			\sym{lbl}_i	&		p{,}φ	&		p{,}φ	&		p{,}φ	&		∅	&		φ	&		p{,}φ	&		φ	&		q{,}φ	&		p{,}φ	&		p{,}φ	&		∅	&		q{,}φ	&		p{,}φ	&		∅	&		q{,}φ	\\
	\sym{maxFst}_i^φ	&		14	&		13	&		12	&	|[unsat]|	-1	&		0	&		1	&		0	&		6	&		7	&		6	&	|[unsat]|	-1	&		0	&		0	&	|[unsat]|	-1	&		0	\\
	\sym{updFst}_i^φ	&		14	&		13	&		12	&		-1	&		0	&		1	&		0	&		6	&		7	&		6	&		-1	&		-1	&		0	&		-1	&			\\
	\sym{sumEff}_i^t	&			&			&		{2{·}6}	&			&			&			&		{-1{·}5}	&			&			&		{2{·}3}	&			&			&			&			&			\\
	\sym{maxLst}_i^φ	&		14	&		1	&		0	&	|[unsat]|	-1	&		5	&		6	&		5	&		6	&		1	&		0	&	|[unsat]|	-1	&		0	&		0	&			&			\\
	\sym{updLst}_i^φ	&		14	&		1	&		0	&		-1	&		5	&		6	&		5	&		6	&		1	&		0	&		-1	&		-1	&		0	&			&			\\	};

\node (l1) at ($(m-4-1.south)!0.5!(m-5-1.north)$) {};
\draw (l1 -| m.west) -- (l1 -| m.east);

\node (l2) at ($(m-10-1.south)!0.5!(m-11-1.north)$) {};
\draw (l2 -| m.west) -- (l2 -| m.east);

  \draw[->, inner sep = 2pt, font={\scriptsize}] (m-1-2) edge (m-1-3)
  (m-1-2) edge["$\binom{1}{0}$"] (m-1-3)
  (m-1-3) edge["$\binom{1}{0}$"] (m-1-4)
  (m-1-4) edge (m-1-5)
  (m-1-5) edge["$_{c{≥}0}$" mygreen, inner sep = 4pt] (m-1-6)
  (m-1-6) edge (m-1-7)
  (m-1-7) edge (m-1-8)
  (m-1-8) edge["$_{d{≥}5}$" mygreen, inner sep = 4pt] (m-1-9)
  (m-1-9) edge (m-1-10)
  (m-1-10) edge["$\binom{1}{0}$"] (m-1-11)
  (m-1-11) edge (m-1-12)
  (m-1-12) edge["$_{d{≥}5}$" mygreen, inner sep = 4pt] (m-1-13)
  (m-1-13) edge (m-1-14)
  (m-1-14) edge (m-1-15)
  (m-1-15) edge["$_{d{≥}5}$" mygreen, inner sep = 4pt] (m-1-16)

  (m-1-4) edge[bend right=80, looseness= 1.3, "$\binom{1}{0}$"'] (m-1-3)
  (m-1-8) edge[bend right=50, "{\color{mygreen}$_{c{≥}0}$} $\binom{-2}{1}$"'] (m-1-6)
  (m-1-11) edge[bend right=80, looseness=1.3, "$\binom{1}{0}$"'] (m-1-10)
  (m-1-16) edge[bend right=80, looseness=1.3] (m-1-15)
  ;

\draw[densely dotted] ([xshift=-1ex,yshift=-0.0ex]m-1-6.south west) |- ([xshift=1ex,yshift=-1ex]m-1-8.south east) -- ++(0,1ex);
\draw[densely dotted] ([xshift=-1ex,yshift=-0.0ex]m-1-2.south west) |- ([xshift=1ex,yshift=-1ex]m-1-4.south east) -- ++(0,1ex);
\draw[densely dotted] ([xshift=-1ex,yshift=-0.0ex]m-1-10.south west) |- ([xshift=1ex,yshift=-1ex]m-1-11.south east) -- ++(0,1ex);
\draw[densely dotted] ([xshift=-1ex,yshift=-0.0ex]m-1-15.south west) |- ([xshift=1ex,yshift=-1ex]m-1-16.south east) -- ++(0,1ex);

\end{tikzpicture}
   \caption{Example of the encoding of the run and path schema from \cref{fig:example-counter-system} with consistent labelling by $φ=\true\Uc{p-¬p≥0}q$.
It demonstrates propagation of counter values and the maximal witness position for $φ$.
Some variables are omitted for conciseness.\label{fig:example-encoding}}
\end{figure}

For convenience, we use not only first-order variables for integer numbers but also boolean, enumeration and natural number types (sorts).
They can, theoretically, be encoded into integers but are more readable and directly supported by, e.g., the z3 SMT solver.
We use notation of the form $\sym{var} : X$ to denote that some variable symbol $\op{var}$ is of some sort $X$.
Mappings with some finite domain $Y$ can be represented by variable vectors of length $|Y|$ that we denote concisely by single variable symbols $\sym{var}:X^Y$.
The shorthand $\ite(\mathrm{cond},\mathrm{prop}, \mathrm{alt})$ represents the \emph{if-then-else} construct.
\Cref{fig:example-encoding} depicts an example of an APS $𝓟$ and its representation in terms of first-order variables and their valuation.
For every state $i∈Q$, we encode the positions of loops in terms of a variable $\sym{typ}_i : \{\sboxminus, ▹, ⊞, ◃\}$ that indicates whether it is outside ($\sboxminus$), inside ($⊞$), the beginning ($▹$), or the end ($◃$) of a loop.
We use $⋄_i$ to abbreviate $\fov{typ}_i=⋄$ for $⋄∈｛⊟, ▹, ⊞, ◃｝$.
The origin is represented by a variable $\sym{org}_i: S$ and the labelling by $\sym{lbl}_i: ｛0,1｝^{\sub(Φ)}$, describing the set $λ_𝓟(i)⊆\sub(Φ)$.
The formula
\[
  \mathrm{fmc}(𝓢,Φ,n) := \mathrm{aps}(𝓢,n) ∧ \mathrm{run}(𝓢,n) ∧ \mathrm{consistency}(n,Φ) ∧ Φ∈\fov{lbl}_0
\]
specifies the shape of $𝓟$, a run and that the initial state is labelled by $Φ$.
The formula components are discussed next.

\subsubsection{Basic structure of APS.}

The basic structure is easily specified as \QPA formula $\mathrm{aps}(𝓢,n)$.
It states that $s_I$ is the origin of the first state ($\sym{org}_0=s_I$), that loops are delimited by $▹$ and $◃$, and that the labelling of states by propositions coincides with that of $𝓢$.

A way to express that the backward transitions from the last to the first state of the loops has a correspondence in $𝓢$ is to build a constraint over all pairs of states from $Q$.
This is, however, quadratic in $n$ and we therefore use a propagation scheme introducing $n$ additional variables $\sym{orgAtEnd}_i : S$.
We let them equal $\sym{org}_i$ where $\sym{typ}_i=◃$ and otherwise be copied from $\sym{orgAtEnd}_{i+1}$, thus propagating backward the origin of the last state of every loop.
The formula
$
 ⋀_{i=0}^{n-1} \,▹_i → ⋁_{(s,μ,Γ,s')∈Δ}\sym{orgAtEnd}_i{=}s ∧ \sym{org}_{i}{=}s'
 $
then guarantees that all backward transitions exist in $𝓢$.
Forward transitions are specified similarly.
We assume a minimal loop length of $2$ due to distinct positions for the first ($▹$) and the last ($◃$) state of each loop but single-state loops can still be represented
(cf.\ \cref{fig:example-encoding}) while increasing the upper bound for the size of path schemas only by one state per loop.
\cref{def:aps} demands that loops be surrounded by identical rows which are not represented explicitly in the encoding.
Instead, runs are required to traverse each representation of a loop at least three times, the first representing the front, the last representing the rear and the remaining representing the actual loop traversals.
The construction distinguishes between the first, second, and last iteration where necessary.

To allow for a simplified presentation, let us assume that there is at most one transition between every two states of $𝓢$, thus being uniquely identified by $\fov{org}_i$ and $\fov{org}_{i+1}$.
The assumption could be eliminated by adding $2n$ additional variables determining explicitly which transition is selected for the represented APS.

\subsubsection{Runs.}
The formula $\mathrm{run}(𝓢,n)$ specifies the shape and constraints of a run in the encoded schema.
Variables $\sym{itr}_i : ℕ$ indicate how often state $i∈Q$ is visited and are thus constraint to equal $1$ outside loops and to stay constant inside each loop.
Infinite iteration of the last loop is represented by the otherwise unused value $0$.

To ensure that the represented run is valid it has to satisfy all the guards at any time.
The variables $\sym{valFst}_i,\sym{valSec}_i,\sym{valLst}_i: ℤ_∞^{C_𝓢}$ hold the counter valuations at state $i∈Q$ when the represented run visits it for the first, the second and the last time, respectively.
Due to flatness each loop is entered and left only once.
Since the guards of the counter system are linear inequalities and the updates are constant, it suffices to check them in the first and last iteration of a loop.
For a term $τ=∑_{j=0}^ℓa_jc_j$ and a variable symbol $\sym{var}:ℤ^{C_𝓢}$ let $τ[\sym{var}]:=∑_{j=0}^ℓ a_j · \sym{var}(c_j)$ denote the substitution of the counter names by the variable symbol (representing the value of) $\sym{var}(c_j)$.
The formula
\begin{multline*}
  \textstyle ⋀_{i=1}^{n-1} ⋀_{(s,μ,Γ,s')∈Δ} \sym{org}_{i-1}=s ∧ \sym{org}_i=s' \\[-0.8ex]
  →
    \textstyle ⋀_{(τ≥b)∈Γ}
      τ[\sym{valFst}_i] ≥ b \ ∧
      (¬ ▹_i → τ[\sym{valLst}_i]≥b)
\end{multline*}
then specifies that the encoded run satisfies the guards whenever taking a forward transition.
Notice that the (forward) transition from state $i-1$ to state $i$ is not taken at the beginning of the last iteration of a loop and thus, its guard must not be checked for the corresponding valuation.
Instead, the guards of the \emph{backward transition} pointing to $i$ must be satisfied from the second iteration on, and are expressed similarly.

It remains to actually specify the counter valuations along the run.
By definition, $\sym{valFst}_0 = 𝟎$.
Outside of loops ($⊟$) we impose
$\sym{valFst}_i = \sym{valSec}_i = \sym{valLst}_i = \sym{valLst}_{i-1} +μ$
where $μ$ is the update of the transition from $i-1$ to $i$.
Inside ($⊞,◃$) we let $\sym{valFst}_i = \sym{valFst}_{i-1} + μ$, $\sym{valSec}_i = \sym{valSec}_{i-1} + μ$ and $\sym{valLst}_i = \sym{valLst}_{i-1} + μ$.
At the beginning ($▹$) of a loop the value in the first iteration is propagated as outside ($\sym{valFst}_i ⩵ \sym{valLst}_{i-1} +μ$), but for the second iteration we impose $\sym{valSec}_i = \sym{valFstAtEnd}_i + μ$ where $μ$ comes from the incoming \emph{backward transition} and is applied to the last value of the previous iteration propagated as above using variables $\sym{valFstAtEnd}_i$.

Having a direct handle on the valuations in the first and second iteration (in terms of the variables $\fov{valFst}_i$ and $\fov{valSec}_i$) as well as the total number of loop iterations ($\fov{itr}_i$), it is tempting to specify the valuations in the last iteration simply by
\[
  \fov{valLst}_i ⩵ \fov{valFst}_i + (\fov{valSec}_i-\fov{valFst}_i)·(\fov{itr}_i-1).
\]
Unfortunately, this formula uses multiplication of variables and hence exceeds Presburger arithmetic.
Instead, the updates over the second to last loop iteration are accumulated in an explicit variable $\fov{lUpd}_i$ such that $\fov{valLst}_i$ can be set to $\fov{valFst}_i + \fov{lUpd}_i$.
We express this accumulation by the formula
\[
  ⋀_{i∈[1,n-2]\atop (s,μ,Γ,s')∈Δ}    \hspace{-0.7em} \left(
  \begin{array}{rcll}
      & (◃_i & ∧\ \fov{org}_{i-1}{=}s ∧ \fov{org}_i{=}s'
        & → \fov{lUpd}_i {=} μ · \fov{itr}_i - μ)\\
    ∧ & (⊞_i & ∧\ \fov{org}_{i-1}{=}s ∧ \fov{org}_i {=}s'
        & → \fov{lUpd}_i {=} μ · \fov{itr}_i - μ + \fov{lUpd}_{i+1})\\
    ∧ & (▹_i & ∧\ \fov{orgAtEnd}_i {=}s ∧ \fov{org}_i {=} s'
        & → \fov{lUpd}_i {=} μ · \fov{itr}_i - μ + \fov{lUpd}_{i+1})
  \end{array}
  \right)
  \begin{array}{@{\hspace{-1.25ex}}l}\\ \\[1.6ex] .\end{array}
\]
Essentially, the multiplication by $\fov{itr}_i$ is distributed over the individual transition updates along the loop.
This is admissible because the individual updates $μ$ appear in the formula not as variables but as constants.
In the formulation above, $\fov{lUpd}_i$ is always zero for states $i$ on the last loop but this is no problem because this particular situation can be handled using $\fov{valFst}_i$ and $\fov{valSec}_i$.
Observe also that the variable $\fov{lUpd}_i$ holds only intermediate results inside and at the end of loops and is undefined outside.
Only for states $i$ that are the beginning of a loop, it holds the precise accumulated loop effect and this value is used for propagation as above.

Using $\fov{lUpd}_i$, the calculation of the valuations in the last iteration of a loop is now specified by $\fov{valLst}_i ⩵ \fov{valFst}_i + \fov{lUpd}_i$.
In the infinitely repeated last loop of the schema, there is no actual last iteration, but the variables are nevertheless used to indicate the limit behaviour by specifying
\[
⋀_{c∈C_𝓢} \quad
   \begin{aligned}[t]
          & (\fov{valFst}_i(c) ⩵ \fov{valSec}_i(c) ⩵ \fov{valLst}_i(c))\\
       & \quad ∨  (\fov{valFst}_i(c) ≫ \fov{valSec}_i(c) ∧ \fov{valLst}_i(c) ⩵ -∞) \\
       & \quad ∨  (\fov{valFst}_i(c) ≪ \fov{valSec}_i(c) ∧ \fov{valLst}_i(c) ⩵ ∞) .
    \end{aligned}
\]

\subsubsection{Consistency.}

The formulae constructed above describe a non-empty augmented path schema in $𝓢$ of which the first state is labelled by $Φ$.
In the following, we develop the components of the formula $\fof{consistency}(𝓢,n,Φ)$ expressing the different cases of \cref{def:consistency}.
Consistency for Boolean combinations (\cref{itm:consistency-bool}) can almost literally be translated to \QPA.
Concerning \cref{itm:consistency-constraint}, constraints of the form $τ≧b$ are not modelled explicitly.
Rather, the formula
\[
  (τ≧b)∈\fov{lbl}_i ↔ τ[\fov{valFst}_i] ≥ b ∧ τ[\fov{valLst}_i] ≥ b
\]
imposes for each $i$ that the represented run satisfies the constraints as if they were guards on all incoming transitions on any state labelled by an atomic constraint.
To express \cref{itm:consistency-X}, variables $\sym{lblAtBeg}_i: 2^{\sub(Φ)}$ propagate labelling information from the start of a loop towards the end.
The condition for formulae $\Xφ∈\sub(Φ)$ is then specified by $(\Xφ∈\fov{lbl}_{n-1} ↔ φ∈\fov{lblAtBeg}_{n-1})$ and for $0≤i≤n-2$ by
\[
    \begin{aligned}[t]
      \fov{ite}\big(\X φ {∈} \fov{lbl}_i,\,
         φ{∈}\fov{lbl}_{i+1} ∧ (◃_i → φ{∈}\fov{lblAtBeg}_i),\,
         φ{∉}\fov{lbl}_{i+1} ∧ (◃_i →φ{∉}\fov{lblAtBeg}_i) \big).
    \end{aligned}
\]

\paragraph{Until \cref{itm:consistency-U-goodlast}.}
Consider a formula $φ= χ \Uc{τ≥b} ψ∈\sub(Φ)$. We first set up some propagations to be able to express \cref{itm:consistency-U-goodlast}.
To access the accumulated value of $τ$ on a single iteration of the last loop we introduce variables $\sym{acc}_i^τ$ for $i∈Q$.
Let the formula $\mathrm{accu}(n,τ)$ be defined as
\[
  \textstyle (\sym{acc}_{n-1}^τ = τ[\sym{lbl}_{n-1}]) ∧ ⋀_{i=0}^{n-2}\, \ite(\sym{itr}_i=0,\, \sym{acc}_i^τ = \sym{acc}_{i+1}^τ + τ[\sym{lbl}_i],\, \sym{acc}_i^τ = \sym{acc}_{i+1}^τ).
\]
It implies that $\sym{acc}_0^τ$ holds the effect of the last loop on the value of $τ$.
\Cref{itm:consistency-U-goodlast} requires that $χ$ holds globally at all reachable states.
For loop states this concerns not only larger states (with respect to $≥$).
The whole loop must be labelled by $χ$.
Using variables $\sym{prpg}_i^χ$ and $\sym{glob}_i^χ$ for $i∈Q$ the formula
\begin{multline*}
    \textstyle \mathrm{glob}(n,χ):= (\sym{prpg}_{n-1}^χ ↔ χ∈\sym{lbl}_{n-1}) ∧ \left(⋀_{i=0}^{n-2} \sym{prpg}_i^χ ↔ \sym{prpg}_{i+1}^χ ∧ χ∈\sym{lbl}_i\right) \\[-1ex]
   \textstyle ∧ (\sym{glob}_0^χ ↔ \sym{prpg}_0^χ) ∧ ⋀_{i=1}^{n-1} \sym{glob}_i^χ ↔ \ite(⊟_i∨▹_i,\, \sym{prpg}_i^χ,\, \sym{glob}_{i-1}^χ)
\end{multline*}
propagates this information through the structure by implying that $\sym{glob}_i^χ$ is true if and only if $χ$ is labelled at all states reachable from $i$.
The information whether $ψ$ holds somewhere on the last loop is made available in terms of the variable $\fov{onLast}^ψ$ by
\[
  \textstyle \fof{fin}(n,ψ) := \fov{onLast}^ψ ↔ ⋁_{i=0}^{n-1} \fov{itr}_i ⩵ 0 ∧ ψ∈\fov{lbl}_i.
\]
Then, condition (\ref{itm:consistency-U-goodlast}) is expressed by
\[
  \mathrm{con\ref{itm:consistency-U-goodlast}}(φ,i) := φ∈\sym{lbl}_i ∧ \sym{acc}_0^τ>0 ∧ \sym{onLast}^ψ ∧ \sym{glob}_i^χ.
\]

\paragraph{Until \cref{itm:consistency-U-counter}.}

\Cref{itm:consistency-U-counter} demands the existence or absence of a witness state proving that $φ= χ \Uc{τ≥b} ψ$ holds.
As before, it would be inefficient to model balance counters and the guards required by the criterion explicitly.
Instead, a formulation is developed that assures that the encoded APS can be assumed to have the necessary counters and guards.
For example, assume some state $i$ is to be labelled by $φ$ and consider the best (maximal) value of the term $τ$ on a path starting at state $i$ and leading to some state satisfying $ψ$, without violating $χ$ in between.
If that value is at least $b$, then there is a state at which a balance counter $c_{τ,i}$ for $i$ and $τ$ would have precisely that value and checking the constraint $c_{τ,i} ≥b$ would succeed.
On the other hand, if the best value is below $b$, then there is no such state.
Even, the dual constraint could be added to any potential witness state and the encoded run would still be valid.

We introduce variables $\fov{maxFst}_i^φ:ℤ_∞$ and $\fov{maxLst}_i^φ:ℤ_∞$ for each $i∈Q$. For the first and last occurrence of state $i$, respectively, they are supposed to hold the maximal value possibly witnessing satisfaction of the constraint, the symbolic value $-∞$ expressing non-existence.
Recall that these positions represent only rows as the first and last iteration of loops represent their front and rear, respectively.
Notice also that the latter value is not defined for positions belonging to the last loop.
Then, \cref{itm:consistency-U-counter} can be expressed for state $i$ in terms of the formula $\fof{con\ref{itm:consistency-U-counter}}(φ,i)$ defined as
\[
        (φ∈\fov{lbl}_i ↔ \fov{maxFst}_i^φ≥b)
      ∧ ((φ∈\fov{lbl}_i ↔ \fov{maxLst}_i^φ≥b) ∨ \fov{itr}_i=0).
\]

\paragraph{Maximal witness.}
The optimal witness value is obtained by a suffix optimum backward propagation from the end to the start of the represented schema. Its \QPA formulation $\fof{witnessMax}(n,φ)$ is comprised of three parts: the
\emph{computation} of the potentially propagated value, the calculation of the accumulated \emph{loop effect} on the value of $τ$ as necessary part of that, and the actual \emph{selection}.
Concerning the selection, the best value is propagated backwards, as long as $χ$ holds.
When the chain breaks, no witness position is properly reachable and the best value is set to $-∞$.
Each state of the schema where $ψ$ holds is a potential witness for preceding states.
Thus, if the propagated value is less than $0$, this state will generally provide a better value for $τ$ than any of its successors.
For example, for the case $χ,ψ∈\fov{lbl}_i$ the formula specifies that $\fov{maxLst}_i^φ = \max(\fov{updLst}^φ_{i},0)$ where variables $\fov{updLst}^φ_{i}$ are assumed to hold the value propagated from state $i+1$.

The overall effect of (all iterations of) a loop on the value of $τ$ is made accessible in terms of variables $\fov{sumEff}^τ_i$ where $i$ is the first state of a loop.
It is obtained by summing up the individual contribution $τ[\fov{lbl}_i] · (\fov{itr}_i-3)$ of each loop state $i$ bound to variables $\fov{eff}^τ_i$.
The effect is multiplied only by $\fov{itr}_i-3$ since the first (front), second (auxiliary), and last (rear) iteration is already accounted for explicitly.
In order to circumvent multiplication of variables in the formula, the variables $\fov{eff}^τ_i$ are themselves defined by distributing the factor ($\fov{itr}_i-3$) over the sum of monomials of the term $τ$.
Assuming $τ$ to have the form $τ=∑_{k=0}^ma_kχ_k$
the loop effect is hence specified by
\begin{multline*}
  \left(\textstyle⋀_{i=1}^{n-2}\ite(▹_i,\,\fov{sumEff}_i^τ {=}\fov{eff}^τ_i,\,  \fov{sumEff}_i^τ {=} \fov{sumEff}_{i-1} + \fov{eff}^τ_i)\right)\\
    \shoveleft{∧ \textstyle ⋀_{i=0}^{n-1}
      \ite(χ_0∈\fov{lbl}_i,\ \fov{eff}^{t,0}_i {=} a_0 · \fov{itr}_i -3a_0,\ \fov{eff}^{t,0}_i {=} 0)}\\
    ∧ \textstyle ⋀_{k=1}^{m} \ite(χ_k∈\fov{lbl}_i,\ \fov{eff}^{t,k}_i  {=} \fov{eff}^{t,k-1}_i + a_k · \fov{itr}_i - 3a_k,\ \fov{eff}^{t,k}_i  {=} \fov{eff}^{t,k-1}_i)
\end{multline*}
where the variables $\fov{eff}_i^τ=\fov{eff}_i^{t,m}$ are to be considered identical.
Then, we can formulate the actual computation of the (potentially) propagated optimum using
\begin{multline*}
    (⊟_i → \fov{updFst}^φ_{i} {=} \fov{updLst}^φ_{i}{=} \fov{maxFst}^φ_{i+1} + τ[\fov{lbl}_i])\\
        ∧ \big(◃_i →
           \fov{updLst}^φ_i {=} \fov{maxFst}^φ_{i+1} + τ[\fov{lbl}_i]
        ∧ \fov{updFst}^φ_{i} {=} \fov{maxAuxAtBeg}^φ_i + τ[\fov{lbl}_i]\\
       \shoveright{∧ \fov{updAux}^φ_{i} {=}  \fov{maxLst}^φ_i + \fov{sumEff}^τ_i\big)}\\
        ∧ \big( ▹_i ∨\ ⊞_i  →
           \fov{updLst}^φ_i {=} \fov{maxLst}^φ_{i+1} + τ[\fov{lbl}_i]
         ∧ \fov{updFst}^φ_i {=} \fov{maxFst}^φ_{i+1} + τ[\fov{lbl}_i] \\
         ∧ \fov{updAux}^φ_i {=} \fov{maxAux}^φ_{i+1} + τ[\fov{lbl}_i] \big).
\end{multline*}
To evaluate \cref{itm:consistency-U-counter-wit-stat,itm:consistency-U-counter-stat} an additional set of auxiliary variables $\fov{maxAux}_i^φ$ and $\fov{updAux}_i^φ$ is used that represents, intuitively, the first real iteration of a loop.
The maximal value is, effectively, propagated through the rear of the loop, then extrapolated over all iterations to the last position on the auxiliary iteration (by adding the accumulated loop effect) and finally through the front row.
Since the value at the last state at the auxiliary iteration depends on that at the first state in the last iteration, the latter is propagated from the beginning to the end of the loop using variables $\fov{maxAuxAtBeg}^φ_i$, similar to the origin above.

Finally, the discussed parts can be combined to express consistency for a formula $χ\Uc{τ≥b}ψ$ by
  \begin{multline*}
    \fof{glob}(n,χ)
        ∧ \fof{accu}(n,t)
        ∧ \fof{fin}(n, ψ)
       ∧ \fof{witnessMax}(n,χ\Uc{τ≥b}ψ)\\
        ∧ \textstyle ⋀_{i=0}^{n-1}
                \fof{con\ref{itm:consistency-U-goodlast}}(χ\Uc{τ≥b}ψ,i)
              ∨ \fof{con\ref{itm:consistency-U-counter}}(χ\Uc{τ≥b}ψ,i).
    \end{multline*}
The structure of the encoding assures that the actual loops are always identically labelled to their front and rear rows.
Thus, by assuring those are consistent, all loops automatically satisfy \cref{itm:consistency-U-unfolding}.
This completes the construction of the formula $\fof{consistency}(𝓢,n,Φ)$ and thereby that of $\fof{fmc}(𝓢,n,Φ)$.

\subsubsection{Properties of the encoding.}
A solution to $\mathrm{fmc}(𝓢,Φ,n)$ yields a $Φ$-consistent APS in $𝓢$ and a run, implying by \cref{thm:correctness} that $𝓢⊧Φ$.
\Cref{cor:existence} implies that if the flat approximation $\FA(𝓢,n)$ contains any run satisfying $Φ$, then $\mathrm{fmc}(𝓢,Φ,2^{p(n)})$ is satisfiable (for a fixed polynomial $p$) at latest.
\begin{theorem}
  \begin{inparaenum}[(i)]
    \item If $\mathrm{fmc}(𝓢,Φ,n)$ is satisfiable, then $𝓢⊧Φ$.
    \item If $\FA(𝓢,n)⊧Φ$, then $\mathrm{fmc}(𝓢,Φ,2^{p(n)})$ is satisfiable.
  \end{inparaenum}
\end{theorem}
The encoding hence provides an effective means to solve the flat model-checking problem based on \QPA satisfiability checking.
A major concern of our construction is to keep the formula as small as possible.
Examining the indexing scheme of variables, we observe that their number is linear in $|Φ|+|𝓢|$ and $n$.
The length of most parts of the formula $\textrm{fmc}(𝓢,Φ,n)$ only depends linearly on $n$ or $n·|Δ|≤n·|𝓢|$.
The parts encoding the guards in $𝓢$ further depend (linearly) on the size of the guard sets associated to the transitions, more precisely, linearly on the total length of all guards.
The components of $\mathrm{consistency}(𝓢,n,Φ)$ are of linear size in $n·|𝓢|$ or $n·|\sub(Φ)|$.
Those concerning atomic constraints and until formulae depend on the length of the constraint terms present in $Φ$.

\begin{theorem}[Formula size\label{thm:formula-size}]
    The  length of $\mathrm{fmc}(𝓢,Φ,n)$ is in $𝓞(n(|𝓢|+|Φ|))$.
\end{theorem}

\section{Evaluation}
\label{sec:implementation}

In order to evaluate whether flat model checking and the \QPA-based encoding can be used to perform verification tasks, we have implemented the procedure and applied it to a set of problems provided by the RERS~Challenge~\cite{DBLP:journals/sttt/HowarIMSBP14}.

The tool \texttt{flat-checker}\footnote{\url{https://github.com/apirogov/flat-checker}} takes a \CLTL specification, a counter system to be verified in DOT format~\cite{DBLP:journals/spe/GansnerN00} and the approximation depth (schema size) and performs the translation of the verification problem to a linear arithmetic formula.
The SMT solver z3~\cite{DBLP:conf/tacas/MouraB08} is used to compute a solution of the formula, if possible, that is subsequently interpreted as satisfying run and presented adequately to the user.
The tool is developed in Haskell and provides a search mode that automatically increases the depth up to a given a bound, in order to potentially find a small witness quickly, before investing computation time in large depths.
A successful search can be continued to find a witness of smallest depth.

The RERS~Challenge~2017\footnote{\url{http://www.rers-challenge.org/2017/}} poses problems as C99 and Java programs that provide output depending on read input symbols and internal state.
The programs have a regular structure but are inconceivable with reasonable effort.
It features a track comprising 100 \LTL formulas to be checked on a program (Problem 1) that is representable as a counter system by treating integer variables as counters.
The counting mechanism of \CLTL admits a more specific formulation of a correctness property, making it more restrictive or permissive than a plain \LTL formula.
For example, a typical pattern in the RERS problem set has the form $¬p\U q$, stating $q$ occurs before $p$.
It can be relaxed to state, e.g., $p$ occurs at most 5 times ($\Fc{p≦5}q$) or less often than $r$ ($\Fc{p-r<0}q$).
A stronger formulation would be that $q$ must occur more often before $p$ ($¬p\Uc{q≧5}q$ or $¬p\Uc{r-q≧5}q$).
To evaluate our procedure on counting properties, we constructed variations of formulae from the \LTL track that express relaxed or strengthened versions of the properties.

By checking negated properties, counterexamples were found at an approximation depth of at most $128$ for all violated formulae, while most formulae could be falsified quickly.
From the original 52 falsifiable \LTL formulae, 43 were falsified after less than 200 seconds per formula at depth at most 64, the remaining 9 took at most 32 minutes per formula and depth 128.
A batch analysis of the whole set of 100 formulae at depth 200 took a total of four days running time (Desktop PC, Intel i5-750 CPU, 4GB RAM).
Some derived \CLTL formulae took significantly longer to be evaluated than the original \LTL formulation.
However, in most cases, the introduction of counting constraints did not increase the evaluation effort significantly.


\section{Conclusion}
\label{sec:conclusion}

The concise representation of runs in terms of augmented path schemas allows for an accelerated evaluation of complex path properties expressed in a powerful specification framework with counting as first-class feature.
We therefore believe that flat approximation provides a promising technique that deserves further investigation.
The underlying theory provides that the procedure is complete on flat systems and, practically, an existing witness will be found eventually unless all of them have an infinitely aperiodic shape.
It can also be used as (incomplete) approach to the satisfiability and synthesis problems of \CLTL.

Although it may eventually hinder problem-specific optimisations, the SMT-based implementation benefits from the engineering effort put into solvers.
 The configurability of, e.g., z3 using specific tactics, provides potential for future improvements.
 It remains to develop and compare different encoding variants.
 Especially, formulations that admit incremental solving could speed up the verification process.
 The primary ambition of our approach is to verify the expressive class of \CLTL properties.
 Our evaluation suggests that this is feasible and, moreover, that flat model checking is well applicable in a general verification context such as the RERS Challenge.

Lifting the theoretical foundation
to linear constraints and counter systems as a class of infinite-state models
is a consequent advancement of the theory of path schemas.
Characterising \CLTL model-checking over flat systems in Presburger arithmetic
fills a gap between corresponding results for temporal logics with and without counting~\cite{DBLP:journals/iandc/DemriDS15,DBLP:conf/rp/DemriDS14,DBLP:conf/concur/DeckerHLST17}.

\subsubsection{Acknowledgement.}
We thank Daniel Thoma for valuable technical discussions.

\bibliographystyle{splncs03}
\bibliography{bibliography}

\newpage
\appendix

\section{Complete Encoding}
\label{apx:encoding-details}

This section presents the details on the formula
\[
  \mathrm{fmc}(𝓢,Φ,n) := \mathrm{aps}(𝓢,n) ∧ \mathrm{run}(𝓢,n) ∧ \mathrm{consistency}(n,Φ) ∧ Φ∈\fov{lbl}_0
\]
introduced in \cref{sec:encoding}.


\subsection{Basic Structure}

The basic structure of APS is specified as \QPA formula
\[
  \fof{aps}(𝓢,n) :=
    \fov{org}_0⩵s_I
    ∧ \fof{typ}(n)
    ∧ \fof{labels}(𝓢,n)
        ∧ \fof{transitions}(𝓢,n).
\]
It states that $s_I$ is the origin of the first state and that loops are delimited by $▹$ and $◃$ in terms of the formula
\[
  \fof{typ}(n) := ⋀_{i∈[1,n-1]} \ite(◃_{i-1} ∨ ⊟_{i-1},\, ⊟_i∨ ▹_i,\, ⊞_i∨◃_i).
\]
To express that the labelling of states by propositions coincides with that of $𝓢$, the formula
\[
\fof{labels}(𝓢,n) := ⋀_{p∈AP, \atop i∈[0,n-1]} p∈\fov{lbl}_i ↔ ⋁_{s∈λ^{-1}(p)}\fov{org}_i ⩵ s
  \]
is used.

\paragraph{Transitions.}
One way to express that each backward transition from the last to the first state of a loop has a correspondence in $𝓢$ is to build a constraint over all pairs of states from $Q$.
This is, however, quadratic in $n$ and we therefore use a propagation scheme introducing $n$ additional variables $\fov{orgAtEnd}_i : S$.
We let them equal $\fov{org}_i$ where $\fov{typ}_i⩵◃$ and otherwise be copied from $\fov{orgAtEnd}_{i+1}$, thus propagating backward the origin of the last state of every loop.
The corresponding formula is
\[\begin{aligned}
  \fof{orgAtEnd}(n) := &\  \fov{orgAtEnd}_{n-1} ⩵ \fov{org}_{n-1} \\
    & ∧  ⋀_{i∈[0,n-2]} \ite(◃_i,\fov{orgAtEnd}_i⩵\fov{org}_i,\fov{orgAtEnd}_i⩵\fov{orgAtEnd}_{i+1}).
\end{aligned}\]
The formula
\[
 \fof{transitionsBwd}(𝓢,n) := ⋀_{i∈[0,n-1]} ▹_i → ⋁_{(s,μ,Γ,s')∈Δ}\fov{orgAtEnd}_i⩵s ∧ \fov{org}_{i}⩵s'
\]
then guarantees that all backward transitions exist in $𝓢$.
Similar propagation chains will be used at other occasions to avoid a quadratic blow-up of the formula size due to information non-locality.
Forward transitions are specified similarly, but without the need for propagation, by
\[
\fof{transitionsFwd}(𝓢,n) := ⋀_{i∈[1,n-1]} ⋁_{(s,μ,Γ,s')∈Δ}\fov{org}_{i-1}⩵s ∧ \fov{org}_i⩵s'.
\]
The combination of the formulae now defines the formula
\[
  \fof{transitions}(𝓢,n) := \fof{orgAtEnd}(n) ∧ \fof{transitionsBwd}(𝓢,n) ∧ \fof{transitionsFwd}(𝓢,n)
\]
used as part of $\fof{aps}(𝓢,n)$ above.
Notice that this encoding assumes a minimal loop length of $2$ due to distinct positions for the first ($▹$) and the last ($◃$) state of each loop.
Single-state loops can still be represented as longer (e.g.\ two-state) loops by combining multiple iterations as one loop that is iterated less often (cf.\ \cref{fig:example-encoding}).
Excluding single-state loops increases the upper bound for the size of path schemas only by one state per loop.

To allow for a simplified presentation, let us assume that there is at most one transition between every two states of $𝓢$, thus being uniquely identified by $\fov{org}_i$ and $\fov{org}_{i+1}$.
The assumption could be eliminated by adding $2n$ additional variables determining explicitly which transition is selected for the represented APS.

\paragraph{Front and Rear Rows.}

The definition of augmented path schemas demands that loops be surrounded by identical rows.
Being identical, these rows are not represented explicitly in the encoding.
Instead, runs will be assumed to traverse each representation of a loop at least three times, the first representing the front, the last representing the rear and the remaining representing the actual loop traversals.
The construction will distinguish between the first, second and last iteration, where necessary.
This is equivalent to representing the states of the front and rear rows individually but allows for a more compact encoding and also provides an efficient way to correlate every loop state to its correspondents on the front and rear.

\subsection{Runs}

The formula $\fof{run}(𝓢,n)$ specifies the shape and constraints of a run in the encoded schema.
It has the form
\[
  \fof{run}(𝓢,n) :=
    \fof{itr}(n)
    ∧ \fof{valuations}(𝓢,n)
    ∧ \fof{guards}(𝓢,n).
\]
Variables $\fov{itr}_i : ℕ$ are used to indicate how often state $i∈Q$ is visited and are thus constraint to equal $1$ outside loops and to stay constant inside each loop.
Since every loop state is used to also represent its counterpart on the front and rear, they are to be repeated at least three times.
Infinite iteration of the last loop is represented by the otherwise unused value $0$.
This is formulated by
\[
  \fof{itr}(n) := \fov{itr}_{n-1}⩵0 ∧ ⋀_{i∈[0,n-2]} (⊟_i ∧ \fov{itr}_i ⩵ 1) ∨ (◃_i ∧ \fov{itr}_i ≫ 2) ∨ \fov{itr}_i ⩵\fov{itr}_{i+1}.
\]
The other components of the formula concern the valuation of counters and evaluation of transition guards.
They are described in the following.

\subsubsection{Counter valuations.}

The valuation of any counter at any position on a run $ρ$ of an encoded APS $𝓟$ is determined unambiguously by the shape of $𝓟$ (in terms of the sequence of states and their origins) and the number of repetitions of every loop.
Yet, in order to formulate that guards need to be satisfied, the counter values will be made explicit in terms of variables $\fov{valFst}_i,\fov{valSec}_i: ℤ^{C_𝓢}$ and $\fov{valLst}_i: ℤ_∞^{C_𝓢}$ for every state $i∈Q$.
They are supposed to hold the counter valuations at the first, second, and last occurrence, respectively, of state $i$ on the represented run.
Naturally, outside loops the first and last valuations are equal and the second does technically not exist, so $\fov{valSec}_i$ does not have a semantically meaningful value.
Nevertheless, all variables are introduced for each state as loops may occur anywhere.
Recall that the states of a loop also represent those of its front and rear rows, so the first and last iteration corresponds to those.
The formula
\[
\begin{array}{r@{}c@{~}l}
\fof{valuations}(𝓢,n) :=  && \fov{valFst}_0 ⩵ 𝟎\\
  &∧& \fof{valRow}(𝓢,n) ∧ \fof{valLoop}(𝓢,n)\\
  &∧& \fof{valFstSecItr}(𝓢,n) ∧ \fof{valLastItr}(𝓢,n)\\
  &∧& \fof{valPropagation}(n) ∧ \fof{loopUpdate}(𝓢,n)
\end{array}
\]
encodes the semantics of counter updates in terms of the valuations in the represented run.
By definition, runs start with the valuation $𝟎$, assigning $0$ to the whole domain.
For row states $i$ (of type $⊟$) the valuation at its (first and only) occurrence is computed from the valuation at the last occurrence of the previous state $i-1$ by adding (elementwise) the update function $μ$ of the transition $(\org(i-1),μ,Γ,\org(i))∈Δ$ from $i-1$ to $i$.
As mentioned earlier, the first and last occurrence are the same and the variable $\fov{valSec}_i$ is deliberately set to equal them as well but could as well be left unconstrained.
Hence, let
\begin{multline*}
  \fof{valRow}(𝓢,n) :=   ⋀_{i∈[1,n-1]} ⊟_i →
  ⋀_{(s,μ,Γ,s')∈Δ} \fov{org}_{i-1}⩵s ∧ \fov{org}_i ⩵ s' → \\
    \fov{valFst}_i ⩵ \fov{valSec}_i ⩵ \fov{valLst}_i ⩵ \fov{valLst}_{i-1} +μ.
\end{multline*}
Recall that we assume that there is at most one transition between every two states in $𝓢$.
Inside ($⊞$) and at the end of loops ($◃$), the counter values are propagated individually for the first, second, and last iteration, expressed by
\begin{multline*}
  \fof{valLoop}(𝓢,n) := ⋀_{i∈[1,n-1]} ⊞_i ∨ ◃_i →\\
  ⋀_{(s,μ,Γ,s')∈Δ} \fov{org}_{i-1}⩵s ∧ \fov{org}_i⩵s' →
    \left(\begin{aligned}
         &\fov{valFst}_i ⩵ \fov{valFst}_{i-1} + μ\\
       ∧\ &\fov{valSec}_i ⩵ \fov{valSec}_{i-1} + μ\\
       ∧\ &\fov{valLst}_i ⩵ \fov{valLst}_{i-1} + μ
    \end{aligned}\right).
\end{multline*}
At the beginning ($▹$) of a loop the value in the first iteration is computed from the preceding position.
The first value in the second iteration is to be computed from the last value of the first iteration.
However, given a state $i$, it cannot be determined \emph{a priori} which state exactly constitutes the end of the loop.
To obtain the value of the state that happens to be the last on the loop, variables $\fov{valFstAtEnd}_i$ are introduced to hold the valuation at the last state during the first iteration throughout the loop and make it thus directly accessible at the beginning.
They are defined using a propagation scheme as above expressed by
\begin{multline*}
  \fof{valPropagation}(n) := \fov{valFstAtEnd}_{n-1} ⩵ \fov{valFst}_{n-1}\ ∧ \\
  ⋀_{i∈[0,n-2]} \ite(◃_i,
    \fov{valFstAtEnd}_i⩵\fov{valFst}_i,
    \fov{valFstAtEnd}_i ⩵ \fov{valFstAtEnd}_{i+1}
  ).
\end{multline*}
Then, $\fov{valSec}_i$ can be set to $\fov{valFstAtEnd}_i + μ$ where $μ$ comes from the incoming \emph{backward} transition of state $i$.
This is specified by
\begin{multline*}
  \fof{valFstSecItr}(𝓢,n) := \\
  ⋀_{i∈[1,n-1]\atop (s,μ,Γ,s')∈Δ}
    ▹_i →
    \left(\begin{aligned}
     (\fov{org}_{i-1}⩵s ∧ \fov{org}_i⩵s'  &→ \fov{valFst}_i ⩵ \fov{valLst}_{i-1} +μ)\\
    ∧ (\fov{orgAtEnd}_i ⩵ s ∧ \fov{org}_i⩵s' &→  \fov{valSec}_i ⩵ \fov{valFstAtEnd}_i + μ)
  \end{aligned}\right).
\end{multline*}

Having a direct handle on the valuations in the first and second iteration (in terms of the variables $\fov{valFst}_i$ and $\fov{valSec}_i$) as well as the total number of loop iterations ($\fov{itr}_i$), it is tempting to specify the valuations in the last iteration simply by
\[
  \fov{valLst}_i ⩵ \fov{valFst}_i + (\fov{valSec}_i-\fov{valFst}_i)·(\fov{itr}_i-1).
\]
Unfortunately, this formula uses multiplication of variables and hence exceeds Presburger arithmetic.
Therefore, we need to specify the value of $(\fov{valSec}_i-\fov{valFst}_i)·(\fov{itr}_i-1)$ differently.
Instead, the updates over the second to last loop iteration are accumulated in an explicit variable $\fov{lUpd}_i$ such that $\fov{valLst}_i$ can be set to $\fov{valFst}_i + \fov{lUpd}_i$.
We express this accumulation by the formula
\begin{multline*}
  \fof{loopUpdate}(𝓢,n) := ⋀_{i∈[1,n-2]} ⋀_{(s,μ,Γ,s')∈Δ}  \\     \begin{array}{rcll}
      & (◃_i & ∧\ \fov{org}_{i-1}⩵s ∧ \fov{org}_i⩵s'
        & → \fov{lUpd}_i ⩵ μ · \fov{itr}_i - μ)\\
    ∧ & (⊞_i & ∧\ \fov{org}_{i-1}⩵s ∧ \fov{org}_i ⩵s'
        & → \fov{lUpd}_i ⩵ μ · \fov{itr}_i - μ + \fov{lUpd}_{i+1})\\
    ∧ & (▹_i & ∧\ \fov{orgAtEnd}_i ⩵s ∧ \fov{org}_i ⩵ s'
        & → \fov{lUpd}_i ⩵ μ · \fov{itr}_i - μ + \fov{lUpd}_{i+1}).
  \end{array}
  \end{multline*}
Essentially, the multiplication by $\fov{itr}_i$ is distributed over the individual transition updates along the loop.
This is admissible because the individual updates $μ$ appear in the formula not as variables but as constants.
Notice that this formulation deliberately multiplies functions with integers, which is to be understood as point-wise application.
Further, the choice of using 0 to mark the infinite iteration of the last loop (as opposed to, e.g., $∞$) is useful here because otherwise the equation would not be well defined, a negative and a positive update could result in having to add $-∞$ and $∞$.
In the formulation above, $\fov{lUpd}_i$ is always zero for states $i$ on the last loop but this is no problem because this particular situation can be handled using $\fov{valFst}_i$ and $\fov{valSec}_i$.
Observe also that the variable $\fov{lUpd}_i$ holds only intermediate results inside and at the end of loops and is undefined outside.
Only for states $i$ that are the beginning of a loop, it holds the precise accumulated loop effect but this suffices since this value is propagated as specified by the formula $\fof{valLoop}(𝓢,n)$ above.

Using $\fov{lUpd}_i$, the calculation of the valuations in the last iteration of a loop is now formulated as
\begin{multline*}
   \fof{valLastItr}(𝓢,n) := \\
   \shoveleft{⋀_{i∈[1,n-1]} ▹_i → ⋀_{(s,μ,Γ,s')∈Δ} \fov{orgAtEnd}_i ⩵ s ∧ \fov{org}_i⩵s' → }\\
    \begin{aligned}
      & \ite \Big( \fov{itr}_i ≫ 0,\ \fov{valLst}_i ⩵ \fov{valFst}_i + \fov{lUpd}_i,
        ⋀_{c∈C_𝓢} (\fov{valFst}_i(c) ⩵ \fov{valSec}_i(c) ⩵ \fov{valLst}_i(c))
    \end{aligned} \quad \\
   \begin{aligned}[t]
       ∨\ & (\fov{valFst}_i(c) ≫ \fov{valSec}_i(c) ∧ \fov{valLst}_i(c) ⩵ -∞) & \\
       ∨\ & (\fov{valFst}_i(c) ≪ \fov{valSec}_i(c) ∧ \fov{valLst}_i(c) ⩵ ∞) & \Big).
    \end{aligned}
\end{multline*}

\subsubsection{Guards.}

To ensure that the represented run is valid it must satisfy all the guards at any time.
The formula $\fof{valuations}(𝓢,n)$ developed above ensures that the variables $\fov{valFst}_i$, $\fov{valSec}_i$, and $\fov{valLst}_i$ faithfully provide the counter valuations when reaching the state $i∈Q$ for the first, the second and the last time, respectively.
Recall that, due to flatness, each loop is entered and left only once.
Since every guard of the counter system is a linear inequality and the effect of the updates of any specific loop is constant, it suffices to check the guard in the first and last traversal in order to guarantee that it is satisfied throughout all repetitions of a particular loop state.

For a constraint term over $C_𝓢$ of the form $τ=∑_{j=0}^ℓa_jc_j$ and a variable symbol $\fov{var}:ℤ^{C_𝓢}$, let $τ[\fov{var}]:=∑_{j=0}^ℓ a_j · \fov{var}(c_j)$ denote the syntactic substitution of the counter names by the variable symbol (representing the value of) $\fov{var}(c_j)$, in analogy to the evaluation of constraint terms using valuations (cf.\ \cref{sec:definitions}).
The formula
\begin{multline*}
\fof{guardsFwd}(𝓢,n) := ⋀_{i∈[1,n-1], \atop (s,μ,Γ,s')∈Δ} \fov{org}_{i-1}⩵s ∧ \fov{org}_i⩵s' → \\
    ⋀_{(τ≧b)∈Γ}
      τ[\fov{valFst}_i] ≧ b
            ∧ (¬▹_i → τ[\fov{valLst}_i]≧b)
\end{multline*}
then specifies that the encoded run satisfies the guards whenever taking a forward transition.
Recall that for some counter $c∈C_𝓢$ the variable $\fov{valLst}_i(c)$ may be assigned a symbolic value.
Thus, a proper interpretation (or expansion) of $≧$ is assumed such that $∞≧b$ holds for every $b∈ℤ$ while $-∞≧b$ holds for none.
Notice that the (forward) transition from state $i-1$ to state $i$ is not taken at the beginning of the last iteration of a loop and thus, its guard must not be checked for the corresponding valuation.
Instead, the guard of the backward transition pointing to $i$ must be verified.
This transition is taken by the encoded run for the first time when entering the second loop iteration.
The guards of backward transitions are thus reflected exhaustively by
\begin{multline*}
  \fof{guardsBwd} := ⋀_{i∈[1,n-1], \atop (s,μ,Γ,s')∈Δ} ▹_i ∧ \fov{orgAtEnd}_i⩵s ∧ \fov{org}_i⩵s' → \\
    ⋀_{(τ≧b)∈Γ} τ[\fov{valSec}_i] ≧ b ∧ τ[\fov{valLst}_i] ≧ b.
  \end{multline*}
Thereby we complete the definition of the formula
\[
  \fof{guards}(𝓢,n) := \fof{guardsFwd}(𝓢,n) ∧ \fof{guardsBwd}(𝓢,n)
\]
and the specification of proper runs in terms of the formula $\fof{run}(𝓢,n)$.

\subsection{Consistency}

The formulae constructed above describe the fact that there is some non-empty augmented path schema in the counter system $𝓢$ of which the first state is labelled by $Φ$.
In the following, we develop the components of the formula
\begin{align*}
  \fof{consistency}(n,Φ) :=
  &\textstyle \hphantom{∧~} ⋀_{(¬φ)∈\sub(Φ)} \fof{consistencyNeg}(n,φ) \\
  &\textstyle ∧ ⋀_{φ∧ψ∈\sub(Φ)} \fof{consistencyAnd}(n,φ,ψ) \\
  &\textstyle ∧ ⋀_{(τ≧b)∈\sub(Φ)} \fof{consistencyCstr}(n,τ,b) \\
  &\textstyle ∧ ⋀_{\Xφ∈\sub(Φ)} \fof{consistencyX}(φ)\\
  &\textstyle ∧ ⋀_{χ\Uc{τ≧b}ψ∈\sub(Φ)} \fof{consistencyU}(n,χ,ψ,τ,b)
\end{align*}
stating that this APS is consistent.
Recall that $Φ$-consistency requires all states of an APS to be consistent with respect to all subformulae of $Φ$.
\Cref{def:consistency} discriminates the structural cases  of a \cLTL formula and therefore the components of the \QPA formulation cover one case each and impose consistency of all states for one subformula of $Φ$ at a time.

\subsubsection{Propositions and Boolean combinations.}
\Cref{itm:consistency-bool} of \cref{def:consistency} can almost literally be translated to \QPA formulae
\begin{align*}
  \fof{consistencyNeg}(n,φ)   & := ⋀_{i∈[0,n-1]} (¬φ)∈\fov{lbl}_i ↔ φ∉\fov{lbl}_i
\intertext{and}
  \fof{consistencyAnd}(n,φ,ψ) & := ⋀_{i∈[0,n-1]}  (φ∧ψ)∈\fov{lbl}_i ↔ φ∈\fov{lbl}_i ∧  ψ∈\fov{lbl}_i.
\end{align*}

\subsubsection{Atomic constraints.}

Concerning \cref{itm:consistency-constraint}, counter guard formulae of the form $τ≧b$ are not modelled explicitly.
Rather, the formula
\[
  \fof{consistencyCstr}(n,τ,b) := ⋀_{i∈[0,n-1]}
    (τ≧b)∈\fov{lbl}_i ↔ τ[\fov{valFst}_i] ≧ b ∧ τ[\fov{valLst}_i] ≧ b.
\]
imposes that the represented run satisfies the constraints as if they were guards on all incoming transitions on any state labelled by an atomic constraint.
Recall that it suffices to assert that the constraint is satisfied at the first and last occurrence of a state.

\subsubsection{Temporal Next.}

To express \cref{itm:consistency-X} of the consistency definition, concerning \emph{temporal next} formulae, variables $\fov{lblAtBeg}_i: 2^{\sub(Φ)}$ are used to propagate labelling information from the first state of a loop forward towards its end.
Similar to the backward propagation of the origin, let
\[
  \fof{propagateX}(n, φ) :=  ⋀_{i∈[1,n-1]} \hspace{-1ex} \ite\big(▹_i,
  \begin{aligned}[t]
    & φ∈\fov{lblAtBeg}_i ↔ φ∈\fov{lbl}_i,\\
    & φ∈\fov{lblAtBeg}_i ↔ φ∈\fov{lblAtBeg}_{i-1} \big).
  \end{aligned}
\]
Notice that it is not necessary to determine the propagation value at the first ($i=0$) state because it is never part of a loop.
The condition is now specified by
\begin{multline*}
    \fof{consistencyX}(n,φ) := \fof{propagateX}(n,φ) ∧ (\Xφ∈\fov{lbl}_{n-1} ↔ φ∈\fov{lblAtBeg}_{n-1}) \\
    ∧ ⋀_{i∈[0,n-2]}
    \begin{aligned}[t]
      \ite\big(\X φ ∈ \fov{lbl}_i,~
        & φ∈\fov{lbl}_{i+1} ∧ (◃_i → φ∈\fov{lblAtBeg}_i),\\
        & φ∉\fov{lbl}_{i+1} ∧ (◃_i →φ∉\fov{lblAtBeg}_i) \big).
    \end{aligned}
\end{multline*}

\subsubsection{Temporal Until: \cref{itm:consistency-U-goodlast}.}

Consider a formula $φ= χ \Uc{τ≧b} ψ∈\sub(Φ)$.
The consistency criterion considers three conditions for until formulae of that form.
Towards defining the corresponding \QPA formula $\fof{consistencyU}(n, χ, ψ,τ,b)$ consider first \cref{itm:consistency-U-goodlast} stating, essentially, that the last loop exhibits a positive effect that eventually proves the formula to hold.
To express the requirements of that condition, the following information is required.
Given a state $i∈[0,n-1]$, first of all, it must be labelled by $φ$ and that information is available in terms of the value of the variable $\fov{lbl}_i$.
Second, assume a variable $\fov{acc}_0^τ: ℤ$ holding the accumulated effect of the last loop on the value of $τ$.
Third, let $\fov{onLast}^ψ:𝔹$ be set to true if and only if $ψ$ occurs as label on some state of the last loop and $\fov{glob}_i^χ:𝔹$ hold if and only if $χ$ holds globally from state $i$ on.
Then, \cref{itm:consistency-U-goodlast} is expressed by
\[
  \fof{con\ref{itm:consistency-U-goodlast}}(φ,i) := φ∈\fov{lbl}_i ∧ \fov{acc}_0^τ≫0 ∧ \fov{onLast}^ψ ∧ \fov{glob}_i^χ.
\]
It remains to formulate the side conditions guaranteeing that the variables actually hold the assumed value.

\paragraph{Accumulated effect of the last loop.}
To describe the accumulated value of $τ$ on a single iteration of the last loop we introduce $\fov{acc}_i^τ$ not only for $i=0$ but for each $i∈[0,n-1]$.
The idea is now to accumulate backwards from $\fov{acc}_{n-1}^τ$ to $\fov{acc}_0^τ$ the effects $τ[\fov{lbl}_i]$ as long as $i$ is part of the last loop (identified by $\fov{itr}_i$ being equal $0$).
Let
\begin{multline*}
\fof{accu}(n,τ) := \\
  \fov{acc}_{n-1}^τ ⩵ τ[\fov{lbl}_{n-1}]
    ∧ ⋀_{i∈[0,n-2]} \ite(\fov{itr}_i⩵0,\ \fov{acc}_i^τ ⩵ \fov{acc}_{i+1}^τ + τ[\fov{lbl}_i],\  \fov{acc}_i^τ ⩵\fov{acc}_{i+1}^τ).
\end{multline*}
It implies, as intended, that $\fov{acc}_0^τ$ holds the effect of the last loop on the value of $τ$.

\paragraph{Reachability of defect- and witness states.}
Consider the evaluation of whether $χ$ holds globally at all \emph{reachable} states.
For loop states $i∈Q$, this means that not only the successors $j≥i$ must be labelled by $χ$ but the whole loop.
Therefore, we employ a propagation scheme with two passes.
First, a backward propagation imposes that variables $\fov{prpg}_i^χ$ hold if and only if all states $j≥i$ are labelled by $χ$.
Based on this information, the intended valuation for $\fov{glob}_i^χ$ is enforced by a forward propagation.
The formula
\begin{multline*}
  \fof{glob}(n,χ):= (\fov{prpg}_{n-1}^χ ↔ χ∈\fov{lbl}_{n-1}) ∧ \left(⋀_{i∈[0,n-2]} \fov{prpg}_i^χ ↔ \fov{prpg}_{i+1}^χ ∧ χ∈\fov{lbl}_i\right) \\
   ∧ (\fov{glob}_0^χ ↔ \fov{prpg}_0^χ) ∧ ⋀_{i∈[1,n-1]} \fov{glob}_i^χ ↔ \ite(⊟_i ∨ ▹_i,\, \fov{prpg}_i^χ,\, \fov{glob}_{i-1}^χ)
\end{multline*}
implies that each variable $\fov{glob}_i^χ$ is true if and only if $χ$ is labelled at all states reachable from $i$.
The information whether $ψ$ holds somewhere on the last loop is made available in terms of the variable $\fov{onLast}^ψ$ by
\[
  \fof{fin}(n,ψ) := \fov{onLast}^ψ ↔ ⋁_{i∈[0,n-1]} \fov{itr}_i ⩵ 0 ∧ ψ∈\fov{lbl}_i.
\]

\subsubsection{Temporal Until: \cref{itm:consistency-U-counter}.}

\Cref{itm:consistency-U-counter} demands the existence or absence of a witness state proving that a formula $φ= χ \Uc{τ≧b} ψ∈\sub(Φ)$ holds.
As before, it would be inefficient to model balance counters and the guards required by the criterion explicitly.
Instead, a formulation is developed that assures that the encoded APS can be assumed to have the necessary counters and guards.

For example, assume some state $i$ is to be labelled by $φ$ and consider the best (maximal) value of the term $τ$ on a path starting at state $i$ and leading to some state satisfying $ψ$, without violating $χ$ in between.
If that value is at least $b$, then there is a state at which a balance counter $c_{τ,i}$ for $τ$ and $i$ would have precisely this value and checking the constraint $c_{τ,i} ≧b$ would succeed.
On the other hand, if the best value is below $b$, then there is no such state.
Even, the dual constraint could be added to any potential witness state and the encoded run would still be valid.

Consider an APS $𝓟$ in $𝓢$ with states $Q=[0,n-1]$ and assume it is consistent with respect to all strict subformulae of $φ$ and admits a run $σ∈\runs(𝓟)$.
Let $x_\mathrm{last}∈ℕ$ be the first position of state $n-1$ on $σ$ and let $\mathsym{maxWit}^{𝓟,σ}_φ: ℕ → ℤ_∞$ denote the discussed function defined for $x∈ℕ$ by
\begin{multline*}
  \mathsym{maxWit}^{𝓟,σ}_φ(x) := \\
    \max(｛⟦τ⟧(\#^{𝓟,σ}_{x,y-1})｜x≤y≤x_\mathrm{last},\ (𝓟,σ,y)⊧ψ,\    ∀_{y'∈[x,y-1]}:(𝓟,σ,y')⊧χ
  ｝∪ ｛-∞｝).
\end{multline*}
We make three essential observations regarding $\mathsym{maxWit}^{𝓟,σ}_φ$.

First, consider the positions $x≤x_\mathrm{last}-|\lastloop(𝓟)|$ preceding the last loop.
For those, $\mathsym{maxWit}^{𝓟,σ}_φ(x)$ accurately determines the maximal value for $τ$ (the symbolic value $-∞$ expressing non-existence of a witness position) unless \cref{itm:consistency-U-goodlast} applies to the state at position $x$ on $σ$.
Assuming that there is a witness position $z≥x_\mathrm{last}$, the last loop must be entirely labelled by $χ$ and if $⟦τ⟧(\#^{𝓟,σ}_{x,z-1})>\mathsym{maxWit}^{𝓟,σ}_φ(x)$, the effect of the final loop on $τ$ must be positive.

Second, if one of \cref{itm:consistency-U-counter-stat,itm:consistency-U-counter-dyn} applies to a (row) state $i∈Q$, then there cannot be a witness position for $φ$ holding at $i$, especially not before $x_\mathrm{last}$, and thus $\mathsym{maxWit}^{𝓟,σ}_φ(x_i) < b$ for the (unique) position $x_i∈ℕ$ of state $i$ on $σ$.
On the other hand, if $\mathsym{maxWit}^{𝓟,σ}_φ(x_i)<b$, then one of \cref{itm:consistency-U-goodlast,itm:consistency-U-counter-stat} applies or any balance counter $c_{τ,i}$ for $i$ and $τ$ would satisfy the guard $c_{τ,i}≪b$ at any witness position $j≥i$ for $φ$.
In the latter case it can thus be assumed that state $i$ obeys  \cref{itm:consistency-U-counter-dyn} in $𝓟$.

Third, a similar point can be made for \cref{itm:consistency-U-counter-wit-stat,itm:consistency-U-counter-wit-dyn} given that $\mathsym{maxWit}^{𝓟,σ}_φ(x_i) ≥ b$.
These conditions imply that there, in fact, is a witness position for $φ$ before the end of the second iteration of the last loop.
Vice versa, the definition of $\mathsym{maxWit}^{𝓟,σ}_φ(x_i)$ demands for some witness position $y≥x_i$.
Then, one of \cref{itm:consistency-U-goodlast,itm:consistency-U-counter-wit-stat} holds or \cref{itm:consistency-U-counter-wit-dyn} can be established without adding extra states.
Hence, $i$ can be assumed to obey one of the conditions in $𝓟$.

Based on these considerations, we introduce variables $\fov{maxFst}_i^φ$ and $\fov{maxLst}_i^φ$ for each state $i∈[0,n-1]$ and \emph{until} formula $φ=χ\Uc{τ≧b}ψ∈\sub(Φ)$ that are supposed represent the value $\mathsym{maxWit}^{𝓟,σ}_φ(x_i)$ at the first position $x_i$ of state $i$ and the value $\mathsym{maxWit}^{𝓟,σ}_φ(x_i')$ at the last position $x_i'$ of $i$, respectively.
Recall that these positions cover only rows as the first and last iteration of loops represent their front and rear, respectively.
Notice also that the latter value is not defined for positions belonging to the last loop.
Then, \cref{itm:consistency-U-counter} is formulated for a state $i$ as
\[
  \fof{con\ref{itm:consistency-U-counter}}(φ,i) :=
  \begin{aligned}[t]
      (φ∈\fov{lbl}_i &↔ \fov{maxFst}_i^φ≧b) \\
      ∧~ ((φ∈\fov{lbl}_i &↔ \fov{maxLst}_i^φ≧b) ∨ \fov{itr}_i⩵0).
  \end{aligned}
\]

\subsubsection{Temporal Until: maximal value to witness.}
The intended value for these variables is specified using a suffix-optimum backward propagation scheme initiated at the end of the represented schema.
We can characterise the values $\mathsym{maxWit}^{𝓟,σ}_φ(x)$ by
\[
  \mathsym{maxWit}^{𝓟,σ}_φ(x_\mathrm{last}) =
    \begin{cases}
      0 & \text{if $ψ∈λ(σ(x_\mathrm{last}))$} \\
      -∞ & \text{otherwise}
    \end{cases}
\]
and for $x∈[0,x_\mathrm{last}-1]$ by
\[
  \mathsym{maxWit}^{𝓟,σ}_φ(x) =
  \begin{cases}
        -∞ & \text{if $χ,ψ∉λ(σ(x))$}\\[1ex]
    0  & \text{if } \begin{aligned}[t]
                      &χ∉λ(σ(x)) \text{ and}\\
                      &ψ∈λ(σ(x))\\[1ex]
                    \end{aligned}\\
    \mathsym{maxWit}^{𝓟,σ}_φ(x+1) + ⟦τ⟧(λ(σ(x)))
      & \text{if } \begin{aligned}[t]
                     &χ∈λ(σ(x)) \text{ and}\\
                     &ψ∉λ(σ(x))\\[1ex]
                   \end{aligned} \\
    \max｛\mathsym{maxWit}^{𝓟,σ}_φ(x+1) + ⟦τ⟧(λ(σ(x))), 0｝& \text{if $χ,ψ∈λ(σ(x))$.}
  \end{cases}
\]
As long as $χ$ holds, the maximal value is propagated backwards.
When the chain breaks at some defect state, no witness position is properly reachable, and the maximal value is set to $-∞$.
Each state of the schema where $ψ$ holds is a potential witness for preceding states.
Thus, if the propagated value at this point is less than $0$, this state will generally provide a better value for $τ$ than any of its successors.
In the \QPA formulation, the above definition is split into the \emph{computation} of the updated value $\mathsym{maxWit}^{𝓟,σ}_φ(x+1) + ⟦τ⟧(λ^{\#}(σ(x)))$ potentially propagated to its predecessor and the actual \emph{selection} of the appropriate value depending on the case.
To express the update across a loop, it is further necessary to express its effect.
These aspects are reflected in the components of the formula
\[
  \fof{witnessMax}(n,φ) := \fof{selectMax}(n,φ) ∧ \fof{calcUpdated}(n,φ) ∧ \fof{loopEffect}(n,τ).
\]

\paragraph{Selection and auxiliary iteration.}
Recall that the encoding does not represent every position of the run and not even every state of the path schema explicitly, namely those situated on loops.
However, the values at the front and rear row of a loop are represented and the propagation scheme hence needs to “jump” from the rear to the front, that is, extrapolate the calculated value over the iterations of the loop.
For that purpose, an additional set of auxiliary variables $\fov{maxAux}_i^φ$ are introduced representing, intuitively, the first actual iteration of a loop---similarly to the variables $\fov{valSec}_i$ above.
Thus, the axillary variables complement the variables $\fov{maxFst}_i^φ$ and $\fov{maxLst}_i^φ$ representing the front and rear rows, respectively.

The case selection is expressed for all three variants by the formula
\begin{multline*}
  \fof{selectMax}(n,φ) := \\
      ⋀_{i∈[0,n-1]}
  \left(\begin{aligned}
      (χ∉\fov{lbl}_i ∧ ψ∉\fov{lbl}_i & → \fov{maxFst}_i⩵\fov{maxAux}_i⩵\fov{maxLst}_i ⩵-∞) \\
    ∧ (χ∉\fov{lbl}_i ∧ ψ∈\fov{lbl}_i & → \fov{maxFst}_i^φ⩵\fov{maxAux}_i⩵\fov{maxLst}_i^φ ⩵ 0) \\
    ∧ (χ∈\fov{lbl}_i ∧ ψ∉\fov{lbl}_i
      & → \left(\begin{aligned}
               & \fov{maxFst}_i^φ⩵ \fov{updFst}_i^φ \\
            ∧~ & \fov{maxAux}_i^φ⩵ \fov{updAux}_i^φ \\
            ∧~ & \fov{maxLst}_i^φ⩵ \fov{updLst}_i^φ
          \end{aligned}\right)) \\
    ∧ (χ∈\fov{lbl}_i ∧ ψ∈\fov{lbl}_i
      & → \left(\begin{aligned}
                & \fov{maxLst}_i^φ ⩵ \max(\fov{updLst}^φ_{i},0)\\
             ∧\ & \fov{maxAux}_i^φ ⩵ \max(\fov{updAux}^φ_{i},0) \\
             ∧\ & \fov{maxFst}_i^φ ⩵ \max(\fov{updFst}^φ_{i},0)
          \end{aligned}\right))
  \end{aligned}\right)
\end{multline*}
where the variables $\fov{updFst}_i^φ$, $\fov{updLst}_i^φ$, and $\fov{updAux}_i^φ$ are assumed to hold the value from the state $i+1$ updated according to the labelling (or the respective initialisation).
For easier reading, expressions of the form $\fov{var1} ⩵ \max(\fov{var2},a)$ are used to abbreviate $\ite(\fov{var2}≫a, \fov{var1} ⩵ \fov{var2}, \fov{var1} ⩵ a)$.

\paragraph{Modelling loop effects.}
The overall effect of (all iterations of) a loop on the value of $τ$ is made accessible in terms of variables $\fov{sumEff}^τ_i$ where $i$ is the first state of a loop.
It is obtained by summing up the individual contribution $τ[\fov{lbl}_i] · (\fov{itr}_i-3)$ of each loop state $i$ bound to variables $\fov{eff}^τ_i$.
The effect is multiplied only by $\fov{itr}_i-3$ since the first (front), second (auxiliary), and last (rear) iteration is already accounted for explicitly.
To circumvent multiplication of variables in the formula, the variables $\fov{eff}^τ_i$ are themselves defined by distributing the factor ($\fov{itr}_i-3$) over the sum of monomials of the term $τ$, as was necessary also for the accumulation of counter updates.
The term is assumed to have the form $τ=∑_{k=0}^ma_kχ_k$
and the effect is hence specified by
\begin{multline*}
  \fof{loopEffect}(n,τ) := \\
  \begin{aligned}
    & \left(⋀_{i∈[1,n-2]} (▹_i → \fov{sumEff}_i^τ ⩵\fov{eff}^τ_i) ∧ (⊞_i ∨ ◃_i → \fov{sumEff}_i^τ ⩵ \fov{sumEff}_{i-1} + \fov{eff}^τ_i)\right)\\
    & ∧ ⋀_{i∈[0,n-1]} \left(\begin{aligned} &  \ite(χ_0∈\fov{lbl}_i,\ \fov{eff}^{τ,0}_i ⩵ a_0 · \fov{itr}_i -3a_0,\ \fov{eff}^{τ,0}_i ⩵ 0)\\
     &∧ ⋀_{k∈[1,m]} \ite(χ_k∈\fov{lbl}_i,\ \fov{eff}^{τ,k}_i  ⩵ \fov{eff}^{τ,k-1}_i + a_k · \fov{itr}_i - 3a_k,\ \fov{eff}^{τ,k}_i  ⩵ \fov{eff}^{τ,k-1}_i)
   \end{aligned}\right)
  \end{aligned}
\end{multline*}
where the variables $\fov{eff}_i^τ=\fov{eff}_i^{τ,m}$ are to be considered identical.

\paragraph{Calculating values to propagate.}
Using the summed-up loop effect, we can now formulate the actual computation of the (potentially) propagated optimum by
\begin{multline*}
  \fof{calcUpdated}(n,φ) :=  \\
  ⋀_{i∈[0,n-2]}
    \left(\begin{aligned}
    (⊟_i & → \fov{updFst}^φ_{i} ⩵ \fov{updLst}^φ_{i}⩵ \fov{maxFst}^φ_{i+1} + τ[\fov{lbl}_i])\\
        ∧ \big(◃_i & →
      \begin{aligned}[t]
           \fov{updLst}^φ_i &⩵ \fov{maxFst}^φ_{i+1} + τ[\fov{lbl}_i] \\
        ∧ \fov{updFst}^φ_{i} &⩵ \fov{maxAuxAtBeg}^φ_i + τ[\fov{lbl}_i]\\
       \operatorname{∧} \fov{updAux}^φ_{i} &⩵  \fov{maxLst}^φ_i + \fov{sumEff}^τ_i\big)
      \end{aligned}\\
        ∧ \big( ▹_i ∨ ⊞_i & →
      \begin{aligned}[t]
          & \fov{updLst}^φ_i ⩵ \fov{maxLst}^φ_{i+1} + τ[\fov{lbl}_i] \\
        & ∧ \fov{updFst}^φ_i ⩵ \fov{maxFst}^φ_{i+1} + τ[\fov{lbl}_i] \\
        & ∧ \fov{updAux}^φ_i ⩵ \fov{maxAux}^φ_{i+1} + τ[\fov{lbl}_i] \big)
      \end{aligned}
    \end{aligned}\right)\\
  \shoveleft{∧ \ite(ψ∈\fov{lbl}_{n-1},\fov{updAux}^φ_{n-1}⩵0, \fov{updAux}^φ_{n-1}⩵-∞)}\\
  \shoveright{∧ \fov{updFst}^φ_{n-1} ⩵ \fov{maxAuxAtBeg}^φ_{n-1} + τ[\fov{lbl}_{n-1}]}\\
\shoveleft{∧ \fov{maxAuxAtBeg}^φ_0 ⩵ \fov{maxAux}^φ_0} \\
∧ ⋀_{i∈[1,n-1]} \ite(▹_i, \fov{maxAuxAtBeg}^φ_i ⩵ \fov{maxAux}^φ_i, \fov{maxAuxAtBeg}^φ_i ⩵ \fov{maxAuxAtBeg}^φ_{i-1})
\end{multline*}
where $φ=χ\Uc{τ≧b}ψ$ is assumed.

The formula $\fof{calcUpdated}(n,φ)$ consists of three parts: the first specifies the updated value, depending on the type of state, the second sets the starting value for the propagation at state $n-1$ for the auxiliary track on which all others depend, and the third makes the value of the auxiliary variables at the begin of each loop available at the corresponding end.

Consider the first part.
Outside of loops (type $⊟$), the first and last encounter of any state fall together, and the updated value is simply calculated from the succeeding position, being the first occurrence of the succeeding state.
A state of type $◃$ marks the end of a loop where the variable $\fov{maxLst}_i^φ$ represent the very last state of its rear row and is hence treated just as other row states.
As mentioned earlier, the auxiliary track can be considered as the first actual iteration of the loop, thus immediately following the front row.
The value of its last state is determined by extrapolating the value at the start of the rear over all iterations by adding the effect of all loop iterations in between.
This may in fact be the correct value of $\mathsym{maxWit}^{𝓟,σ}_φ$ at this point.
However, in case there is a defect on the loop or the effect of the loop is negative, the witness assumed by the extrapolation is not reachable without violating $χ$ in between or may not provide the maximal value for $τ$, respectively.
Nevertheless, as the value is passed along it traverses all positions of the loop.
Then, if the loop does have a defect, the selection determined by the formula $\fof{selectMax}$ would necessarily reset that value to either $0$ or $-∞$ and provide a correct value from that point on.
Similarly, if the overall effect of the loop is negative and there is a witness providing a higher value of $τ$, this witness would be found on the first iteration and the selection would again promote this one as soon as it is encountered.
Hence, upon reaching the first state of the loop, the propagated value on the auxiliary track is in fact correct.
It is transferred back to the end of the loop by the last part of the formula (by variables $\fov{maxAuxAtBeg}_i^φ$) and then used to correctly determine the value of $\fov{maxFst}_i^φ$.

Therefore, assessing consistency \cref{itm:consistency-U-counter} as stated by the formula $\fof{con\ref{itm:consistency-U-counter}}$ is appropriate, at least for those states to which \cref{itm:consistency-U-goodlast} does not apply.
Note that, if the latter does apply to  some state, the evaluation of the other criterion is irrelevant.

\subsubsection{Temporal Until: consistency.}

Based on the developments above, the cases for the consistency criterion are combined to express consistency for temporal until formulae $χ\Uc{τ≧b}ψ$ by
\[
  \fof{consistencyU}(n, χ, ψ, τ,b) := \begin{aligned}[t]
      & \ \fof{glob}(n,χ)
        ∧ \fof{accu}(n,τ)
        ∧ \fof{fin}(n, ψ)\\
       ∧ &\ \fof{witnessMax}(n,χ\Uc{τ≧b}ψ)\\
        ∧ & ⋀_{i∈[0,n-1]}
                \fof{con\ref{itm:consistency-U-goodlast}}(χ\Uc{τ≧b}ψ,i)
              ∨ \fof{con\ref{itm:consistency-U-counter}}(χ\Uc{τ≧b}ψ,i).
    \end{aligned}
\]
The structure of the encoding assures that the actual loops are always identically labelled to their front and rear rows.
Thus, assuring those are consistent, all loops automatically satisfy \cref{itm:consistency-U-unfolding}.

This completes the construction of the formula $\fof{consistency}(𝓢,n,Φ)$ and thereby that of $\fof{fmc}(𝓢,n,Φ)$.

\end{document}